\documentclass[preprint,superscriptaddress,showpacs,amsmath,amssymb,aps,prx,floatfix]{revtex4-1}
\usepackage[utf8]{inputenc}
\usepackage{amsmath}
\usepackage{xcolor}
\usepackage{amsfonts}
\usepackage{amssymb}
\usepackage[normalem]{ulem}
\usepackage{graphicx}
\usepackage{csvsimple}
\usepackage{hyperref}
\usepackage{nameref}
\usepackage{longtable}
\usepackage{booktabs}
\usepackage{soul}
\usepackage{xr}
\usepackage{setspace}
\makeatletter
\usepackage[margin=1in]{geometry}
\newcommand*{\addFileDependency}[1]{
  \typeout{(#1)}
  \@addtofilelist{#1}
  \IfFileExists{#1}{}{\typeout{No file #1.}}
}
\makeatother

\newcommand*{\myexternaldocument}[1]{
    \externaldocument{#1}
    \addFileDependency{#1.tex}
    \addFileDependency{#1.aux}
}

\myexternaldocument{SI}

\begin{document}

\title{Don't follow the leader: Independent thinkers create scientific innovation}


\author{Sean Kelty}
    \affiliation{Department of Physics \& Astronomy, University of Rochester, Rochester, NY 14607, USA}
   \author{Raiyan Abdul Baten}
    \affiliation{Department of Computer Science, University of Rochester, Rochester, NY 14607, USA}
\author{Adiba Mahbub Proma}
     \affiliation{Department of Computer Science, University of Rochester, Rochester, NY 14607, USA}
\author{Ehsan Hoque}
	 \affiliation{Department of Computer Science, University of Rochester, Rochester, NY 14607, USA}
\author{Johan Bollen}
    \affiliation{Luddy School of Informatics, Computing and Engineering, 919 E. 10th St., Bloomington, IN 47408, USA}
   \author{Gourab Ghoshal}
    \email[Correspondence email address: ]{gghoshal@pas.rochester.edu}
    \affiliation{Department of Physics \& Astronomy, University of Rochester, Rochester, NY 14607, USA}
    \affiliation{Department of Computer Science, University of Rochester, Rochester, NY 14607, USA}

\begin{abstract}
\begin{singlespace}
Academic success is distributed unequally; a few top scientists receive the bulk of attention, citations, and resources. However, do these ``superstars" foster leadership in scientific innovation? We introduce three information-theoretic measures that quantify novelty, innovation, and impact from scholarly citation networks, and compare the scholarly output of scientists who are either not connected or strongly connected to superstar scientists. We find that while connected scientists do indeed publish more, garner more citations, and produce more diverse content, this comes at a cost of lower innovation and higher redundancy of ideas. Further, once one removes papers co-authored with superstars, the academic output of these connected scientists diminishes. In contrast, authors that produce innovative content without the benefit of collaborations with scientific superstars produce papers that connect a greater diversity of concepts, publish more, and have comparable citation rates, once one controls for transferred prestige of superstars. On balance, our results indicate that academia pays a price by focusing attention and resources on superstars. 
\end{singlespace}
\end{abstract}

\maketitle

    \section{Introduction}
    
    ``To truly make an apple pie from scratch you must first invent the universe"---a quote attributed to Carl Sagan~\cite{Cliff_2021}---illustrates the idea that the process by which individuals create is contingent upon the elements on which that creation is based. Whether creating a new piece of music, going about daily routines, or engaging in scientific research, people's actions are founded in the information, experiences, and relationships that they have establish by themselves and through others~\cite{McAndrew:2014yi,Muller:2019ny,Hazarie:2020zb,Chen:2022tt}. Each person has their own basis of knowledge that stems from their own lived experiences while also existing in a network of relationships through which they share experiences and knowledge with each other, thereby informing a collective understanding among a network of connected individuals~\cite{Rodriguez_2016}. Within such networks, hierarchies can emerge in which some actors exert greater social influence over the network and thus the creative process that it supports, while others may influence only those closest to them or no one at all~\cite{Holme:2006bf}. This social hierarchy is common in the societal dynamics of government and politics, where some individuals and institutions exert a great degree of influence over the flow of information in the system and opinion formation~\cite{Ghoshal:2007la,Recuero:2019jp,Dubois:2014ch}. 


Academia is not immune from the emergence of social hierarchies; some academics can function as figures of authority due to the merit and influence of their work and their prominent position in a network of academic collaborations. Citations as an indicator of academic influence \cite{RADICCHI2017704} have long been known to be distributed very unequally\cite{hindex:hirsch2005}, with a minority of a few scientists receiving most citations. Such inequality may be increasing at a global level\cite{global-citation-inequality}, at least with respect to citation numbers. In academic publishing, biasing effects like this have been studied under the lens of the Matthew Effect, where success begets more success and early success compounds into a cumulative advantage as the ``rich get richer" \cite{Matthew-Effect}. There are arguments that this effect is beneficial for academia; the rewards of top researchers are proportional to their contributions, which ensures the ``epistemic security'' of the field~\cite{encyclopedia-creativity}. This thinking is aligned with the notion that science should operate as a meritocracy; those who contribute the most are also valued the most, and will therefore be most influential. Indeed, there is a high degree of trust in our most successful academics and the value of their mentorship. For instance, junior researchers collaborating with top scientists at the early stages of their career are likely to become top-cited scientists themselves, especially those at less prestigious universities~\cite{top-scientist-collab-success}. Inexperienced academics can benefit from apprenticeships with top scientists; the ``chaperoning" of early-career scientists leads to higher rates of publication in high-impact journals \cite{chaperone-effect}. These relationships are frequently mutually beneficial. Less visible authors benefit from more opportunities to publish papers in high quality journals that attract larger audiences, whereas top scientists gain collaborators with unique skills to produce more high quality work~\cite{coauthorship-top-ordinary}. Close collaboration of less visible academics with those in the upper echelons can furthermore create opportunities for a first-mover advantage, inducing a positive feedback loop and early bandwagoning of innovative ideas~\cite{Abrahamson-innovation-diffusion}.

While top academics (sometimes referred to as ``superstars") may make consistent and high impact contributions that benefit their field and collaborators, their status as superstars may also have deleterious effects due to the subsequent concentration of resources and attention. For instance, it has been shown that the collaborators of academic superstars experience a 5 to 9$\%$ drop in publication rates after the sudden death of that superstar~\cite{Superstar-Extinction}, highlighting their dependence on the superstar's collaboration. In fact, it is unclear whether collaborating with superstars truly fosters independent career development~\cite{Clauset:eu,Janosov:2020ri} 
Furthermore, superstars can induce a high degree of inequality in the distribution of research funding due to a funding Matthew-effect. Those who receive funding accumulate twice as much research funding afterwards compared to those who submitted similarly valued proposals but found themselves, by chance, just below the funding threshold. There is no evidence that this accumulation of research funding is due to actual achievements enabled by previous funding~\cite{Matthew-Effect-Science-Funding,Matthew-Effect-Model}. If successful collaborations with superstars lead to early funding success, this can induce a superstar-fueled funding cycle that increasingly widens the gap between scientific haves and have-nots. 


The topology, structure, and characteristics of scientific collaboration networks may play an important role in these effects since they shape both the production and dissemination of ideas, potentially with conflicting outcomes. Tightly connected networks could be more efficient in distributing and leveraging knowledge thereby yielding higher productivity, but may at the same time lead to a decline of diversity, reducing exploration and discovery \cite{exploitation-exploration-Lazer,Rodan-Heterogeneity-Innovation,Chang-knowledge-diffusion-imitation-innovation}. Although some spillover effects may occur, i.e.~collaborators of highly-acclaimed authors benefit by proxy~\cite{Novelty-Recognition-MatthewEffect}, it is not clear whether the concentration of attention of resources towards superstars yields more novel and innovative research. This is a particularly relevant issue with the rise of interdisciplinary research which relies on the ability of scientists to collaborate in equitable teams that foster creativity and innovation across various research fields \cite{flat:evans2022}.

To investigate the effects of superstar influence on academic productivity, impact, and innovation, we perform a comprehensive analysis of the American Physical Society corpus. Following~\cite{Superstar-Extinction}, we define superstars as academics who are among the top .1\% in terms of their h-index~\cite{h-index-justification,h-index}.  We extract the semantic content of over 250,000 abstracts, defining a number of information-theoretic measures to quantify the novelty and innovation of each paper. We augment this with analysis of publication and citation rates, and examine the difference in academic output between researchers who collaborate with or cite frequently papers by superstars against those with little-to-no connection to such superstars. We find that at the individual level, collaborators and frequent citers of superstars, publish more, garner higher citations and produce papers with more diverse content compared to other academics. However, their work is no more innovative than the rest of the corpus and its content is more redundant. Further, once one excludes papers co-authored with superstars, their publication and citation output are no different from the rest of the corpus and in some cases output is lower.

Focusing on early career researchers, we find that those who frequently collaborate with superstars in the beginning of their careers, do eventually go on to produce impressive academic output, although once the collaboration is removed, their output in terms of publication rates, citation impact, and innovation is significantly diminished. On the other hand, early career researchers that produce innovative content without the benefit of early superstar collaboration, continue to produce such content over the rest of their careers. They publish more then early collaborators of superstars and accrue similar citation numbers, once one controls for the collaboration itself.

\section{Results}

\subsection{Data}

We use the American Physical Society (APS) corpus \cite{APS-dataset} that contains articles published in APS journals since 1893. The data set contains full citation data, i.e.~the citations pointing from the references of one article to another, allowing a reconstruction of the full citation network among all articles, including article-specific fields such as DOI, journal, volume, issue, first page and last page OR article id and number of pages, title, authors, affiliations, publication history, PACS codes, table of contents heading, article type, and copyright information. Given that the data does not include article abstracts, we used a web-scraping algorithm~\cite{beautiful-soup} to collect abstracts for 250,628 articles corresponding to between 35-40\% of all published papers across the different APS journals (Fig.~S1). We note that around ~1\% of these articles have references not contained in the APS citation network, and on average we scraped abstracts for 38\% of paper references. The distribution of citations and h-index are both heavy-tailed (Fig.~S2), with the average number of citations being 14.4 and the average h-index 1.74. Author disambiguation was done using a rule-based scoring method~\cite{author-disambiguation} (Cf. Sec.S1.2) We consider authors who first publish on or after 1970, and define superstars as those with the top .1\% of h-index in the corpus, corresponding to an h-index threshold of 21. This yields 303 superstars among 292,394 authors. The summary statistics can be found in Tab.~S1.
 
In order to extract topics from the collected abstracts, we use an unsupervised Latent Dirichlet Allocation (LDA) algorithm on phrases (P-LDA) \cite{PLDA} to establish vector embeddings for phrases and documents within our corpus. Stop words in the corpus were removed, all words were lemmatized, and phrases were determined based on a significance score that determined whether or not phrases occurred due to random chance. These vector embeddings have dimensionality $k$ correspoding to the number of topics defined for our corpus. P-LDA utilizes Gibbs Sampling to generate distributions of topics over phrases as well as documents~\cite{Lee:2022tf}, from which novelty scores can be extracted based on topic-spread. We choose a number of topics $k$ based on the UMass coherence measure (\cite{UMASS-Coherence}), the value of which first stabilizes at $k=25$ topics (Fig.~S3). Tab. ~S2 shows the top 10 terms per topic. The resulting output for each document $u$ is a $k$-dimensional vector ${\bf v}^u$ whose elements $v^u _i$ correspond to the frequency of topic $i$ extracted from its abstract (example in Tab.~S3).   
       
\subsection{Novelty, innovation and redundancy}

Novelty detection in the literature has been implemented in a variety of ways~\cite{Novelty-Detection-Review}, such as contextualizing novelty in machine learning as information retrieval~\cite{Novelty-Detecion-TREC,Ghoshal-Novelty-Detection}, distant combinations of ideas via citation relations~\cite{Uzzi-Atypical-Recomb}, first-pass combinations of concepts never before connected~\cite{Schumpeter:1982}, knowledge-graphs of concepts within social networks~\cite{Rodan-Heterogeneity-Innovation}, and agent-based simulations of social and individual learning~\cite{Chang-knowledge-diffusion-imitation-innovation}.

Here we rely on document-level embeddings that represent a distribution of all topics contained within the abstract of given paper, using which one can define the topic diversity in terms of a paper, its references, and articles that cite the paper. Using this, we define a variety of metrics capturing different aspects of novelty and innovation.  

Coupling connections between authors and the content of their works can then elucidate the influence that superstars have on the success of and novelty produced by other academics.

 \noindent {\it Entropy:}
For a given document $u$, we define the Shannon entropy as   
\begin{equation}
    I^{(S)}_u= -\sum^{k}_{i=1} v^u_i \ln v^u_i,
    \label{eq:shannon}
    \end{equation}
 The expression quantifies the average level of ``surprise" or uncertainty over the outcomes of a random variable~\cite{Information-Theory-Book}. In this context, papers focusing on limited number of topics in abstracts will yield low values of $ I^{(S)}_u$, whereas those with a wide diversity of topics will yield a larger value of the entropy.
 
 \noindent {\it Reference and Citation Diversity:}
While $I^{(S)}_u $ measures the ``surprise'' with respect to a paper's content, in this case its abstract, references and citations refer to the degree that the ideas in a given paper were inspired by other papers (references) or of inspiration to other papers (citations). We can thus measure the novelty of a paper, or its Information Diversity \cite{Aral-Dhillon}, by evaluating the dispersion of the topics of its references or the citations its receives. The greater the variance of the topic distribution, the higher the information diversity. For a set $X^{u}$, that can represent either the references in paper $u$, or citations to paper $u$, we define the quantity,
\begin{equation}
I^{(X)}_u = \frac{1}{|X^{u}|} \sum_{l\in X^{u}} \left[1-\cos\left({\bf v}^l,\overline{X^{u}}\right)\right]
\label{eq:refcite}
\end{equation}
where $\cos\left({\bf v}^l,\overline{X^{u}}\right)$ is the cosine similarity of the vector embedding of a particular reference/citation ${\bf v}^l$ with the average over the vector embeddings of all references/citations in the set $X^{u}$. We can as such define \emph{reference diversity} and \emph{citation diversity} as the information diversity over the references from a paper and citations to the paper respectively.

\noindent {\it Innovation:}
The metrics defined thus far are based on topic models expressed as topic distributions per document derived from the words in their content (abstracts). These metrics capture topic diversity of the paper itself, or its influences, but does not express the degree to which the paper expanded the literature through innovation. In other words, they express what document themselves are about, but not whether this adds to the diversity of the literature. We therefore define Innovation as the degree to which the document adds topics  in new combination to the literature \cite{Kuhn1962, diversity-innovation-paradox}. Specifically, innovation in this context, is a measurement of when terms were first introduced or combined in the corpus (cf. Sec.~S1.4 and Fig. S4). Coupled with the novelty measures, this allows us to track how the diversity of ideas correlates with new conceptual recombinations and co-occurrences of terms. Following this logic, we define the Innovativeness of paper $u$ as 
\begin{equation}
    I_u^{(I)} = \frac{1}{2} \sum_{w_1\neq w_2 \in u} \mathcal{I}(w_1,w_2;u)
    \label{eq:innovate}
\end{equation}
where $w_1$ and $w_2$ are distinct terms in paper $u$, $\mathcal{I}(w_1,w_2;u)$ is an indicator function that is $1$ if terms $w_1$ and $w_2$ are first seen within the corpus in paper $u$ and 0 otherwise, and the $\frac{1}{2}$ prefix accounts for double counting. To remove spurious conceptual links due to chance or extreme rarity, we calculate a point-wise mutual information for all links as the log ratio of co-occurrence probability over the individual probabilities of each concept \cite{diversity-innovation-paradox}. In Fig.~S5 we determine the Pearson's $r$ correlation coefficients between each measure and  find only weak correlations, indicating that each measure captures a different aspect of academic output. 

\noindent {\it Redundancy:}
Finally, in a related context, in the field of creative ideation, it has been reported that inspirees stimulated by highly creative alters, tend to generate more creative ideas~\cite{Baten-divergent-thinking, Baten_2021,Baten_2022}.  However, as a group, the inspirees ideas was found to be similar to each other leading to redundancy  in generated ideas over time at the group level. To check whether a similar effect manifests in academic publishing, we compute the cosine similarity score between papers $u,u'$ in the set $P(G,s,t)$ thus
\begin{equation}
\textrm{Sim}(G,s,t) = \frac{2}{\left|P(G,s,t)\right|(\left|P(G,s,t)\right| - 1)} \sum_{u,u'\in P(G,s,t)} \cos({\bf v}^u, {\bf v}^{u'}).
\label{eq:sim}
\end{equation}

\subsection{Superstar statistics}

\begin{figure}[t!]
\includegraphics[width=0.75\linewidth]{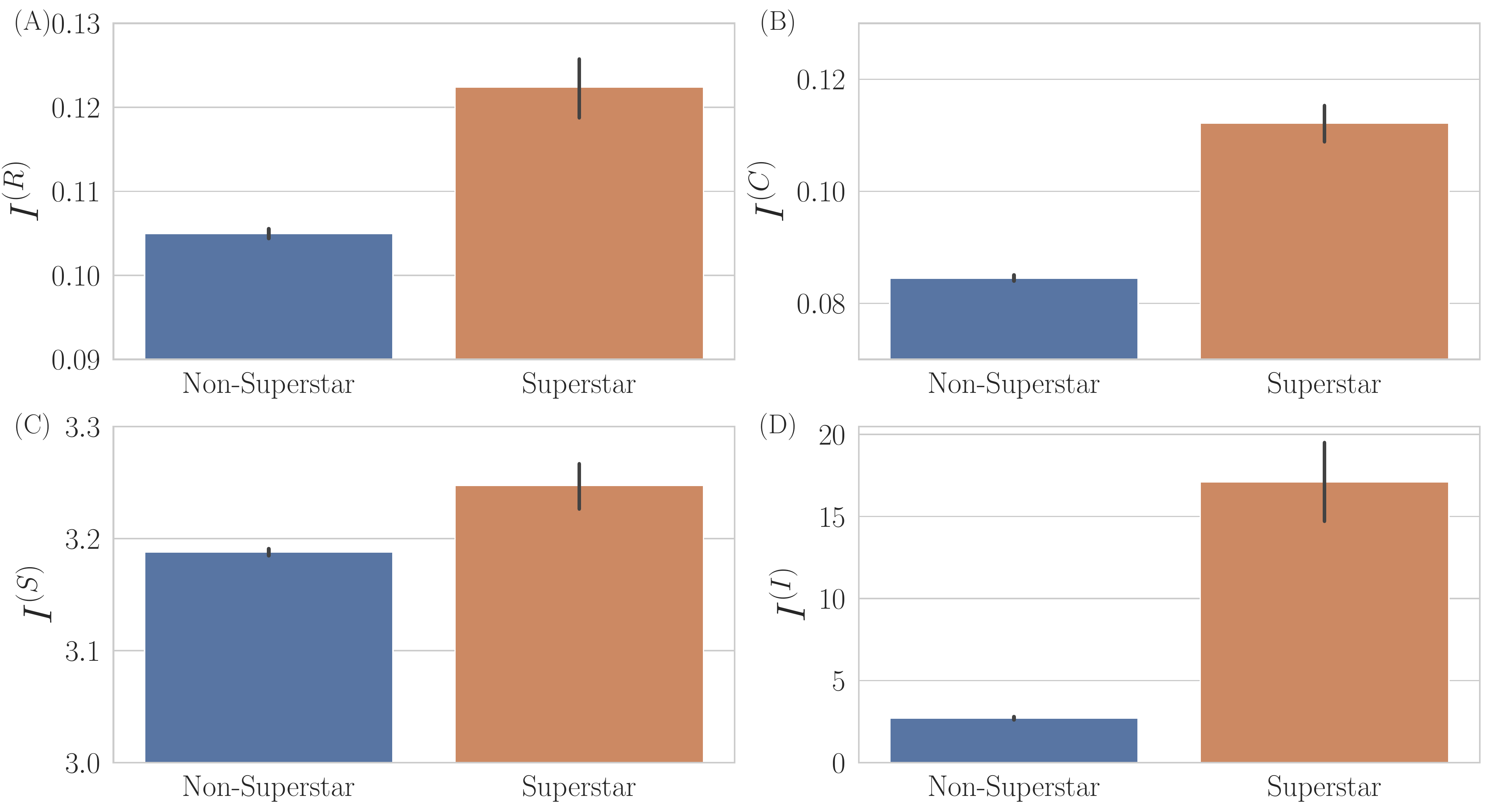}
\caption{{\bf Average author-level statistics of novelty and innovation} \textbf{A} Reference Diversity, \textbf{B} Citation Diversity, \textbf{C} Shannon Entropy, \textbf{D} Innovation. The orange bar is for superstars (h-index $\ge$ 21) and the blue bars correspond to all other authors in the corpus.}
\label{fig:author_ss_nov_stats}
\end{figure}

We next examine whether the novelty and innovation produced by superstars are significantly different from the rest of the academic corpus. In Fig.~\ref{fig:author_ss_nov_stats} we plot the Reference and Citation diversity (Eq.~\eqref{eq:refcite}), the Shannon entropy (Eq.~\eqref{eq:shannon}) and Innovation (Eq.~\eqref{eq:innovate}) comparing the set of superstar academics against the rest of the authors in the corpus. In terms of reference diversity, citation diversity and Shannon entropy, superstars outperform the remaining academics by $20\%$, $15\%$, and $2\%$ respectively. That is, superstars are inspired by a higher diversity of content, publish works that are more conceptually diverse, and inspire a wider array of publications than non-superstars. The starkest contrast can be seen in terms of Innovation, where there is a factor of ten difference between superstars and other academics indicating that the former are more prolific in introducing new combinations of terms. We note that there is a monotonic dependence of the metrics with number of publications for all academics, although the effect is more pronounced for superstars (Fig.~S6). Furthermore, there is also a monotonic dependence of citations received by a paper $u$ and the novelty/innovation metrics (once again more pronounced for superstars) indicating that an increase in conceptual diversity and the ability to connect concepts for the first time is rewarded in terms of more attention paid to that paper (Fig.~S7).

\subsection{Superstar influence}
Having established that superstars outperform other academics in terms of our metrics, we next determine to what degree superstars affect the academic output of their collaborators and their ``inspirees" (those inspired by their work). Inspirees are authors that cite a superstar's papers, for whom we determine the degree of inspiration by the frequency of citations. 
We examine inspirees both at the group- and individual-levels. At the group-level, we center the superstar in a network of inspirees where the degree of inspiration is the number of times a researcher cites the superstar. 
We then partition the inspirees into groups based on their degree of inspiration, where the upper bounds for each bin are the top 10$\%$ of inspirees, 20$\%$, 30$\%$, 50$\%$, and 100$\%$. These groups represent increasingly weakening ties to a given superstar; those in the top 10 percent are the most actively inspired, while the bottom 50 percent typically cite the superstar only once. Note that some inspirees in the bottom 50 group of one superstar may be in the top group of another superstar. The increasing bin sizes are chosen to account for the decreasing frequency of inspired citations among the least-inspired inspirees, such that there are sufficient number of papers compared between groups.

\begin{figure}[t!]
\centering
\includegraphics[width=.95\linewidth]{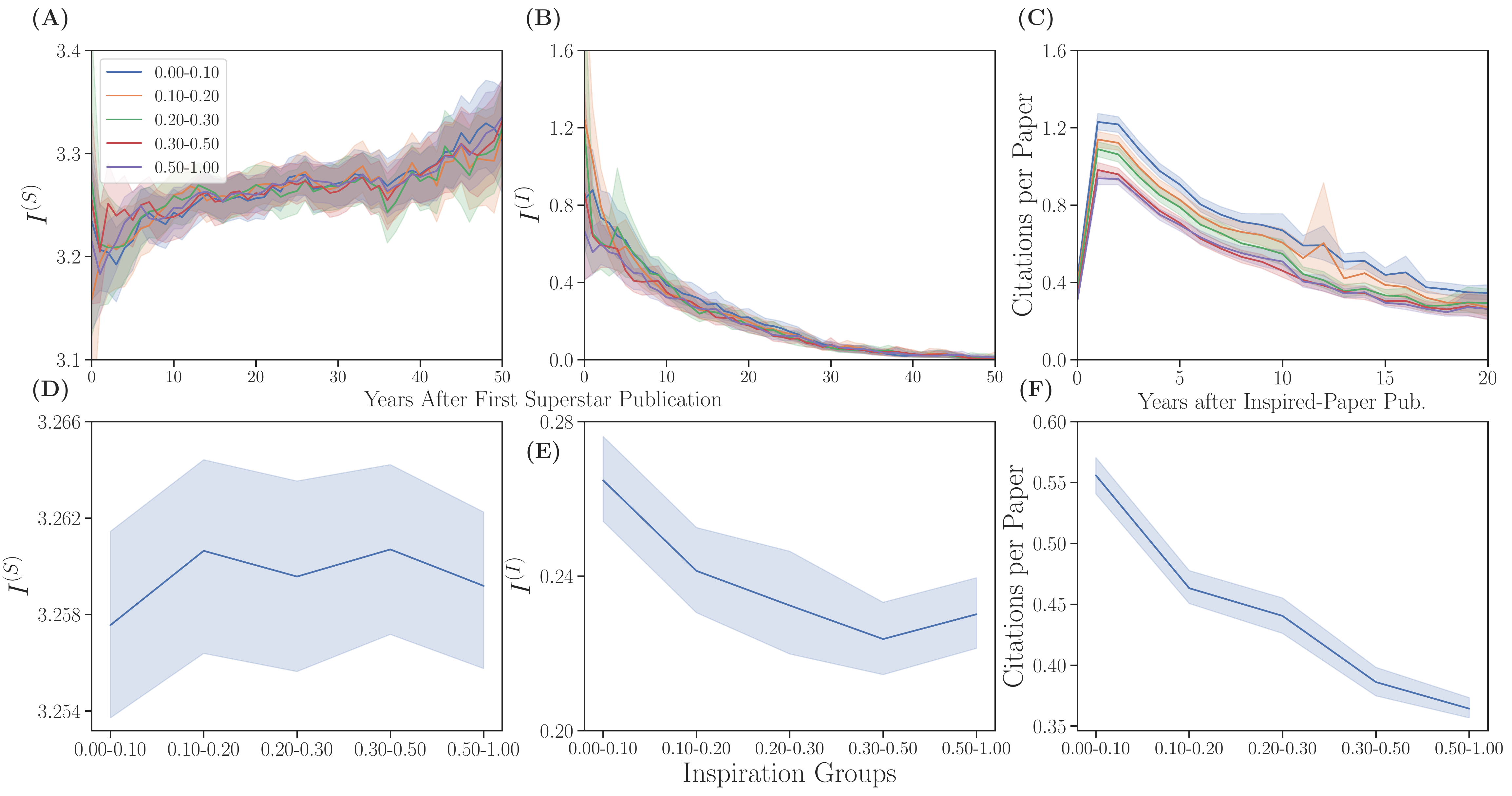}
\caption{{\bf Novelty and Innovation statistics at the group-level} Temporal trajectory of average paper-level statistics. \textbf{A}: Shannon Entropy, \textbf{B}: Innovation, \textbf{C}: Citations per-paper. Aggregated group-level statistics \textbf{D}: Shannon Entropy, \textbf{E}: Innovation, \textbf{F}: Citations per-paper. Curves indicate averages, shaded area 95\% confidence interval. }
\label{fig:group_nov_statistics}
\end{figure}

Given that we are interested in the temporal evolution of superstar influence on the novelty and innovation of the inspirees, we denote the year of the first superstar publication as $t_0 = 0$ and for every susbsequent year $t > t_0$, we consider the set of publications by the inspirees who cite the superstar. For each partitioned group, we calculate the average novelty of all of the publications in year $t$ per partition. Denoting the set of papers inspired by superstar $s$ for partition $G$ at year $t$ as $P(G,s,t)$, the average novelty scores are computed as     
\begin{equation}
        \langle I_u ^{(l)}\rangle_{G,s,t} = \frac{1}{|P(G,s,t)|} \sum_{u\in P(G,s,t)} I_u^{(l)}
\label{eq:avgnov}      
\end{equation}      
 where $l = S, X, I$ is the novelty or innovation score of paper $u$.  
             
We plot the results of our analysis in Fig.~\ref{fig:group_nov_statistics}. In terms of the temporal evolution of the Shannon entropy, while there is a monotonic increase---reflecting an increase in the body of knowledge with time (Fig.~S8)---we find little-to-no differences across the groups as seen in Fig.~\ref{fig:group_nov_statistics}{\bf A}. Averaging over the entire temporal range also indicates a flat trend (Fig.~\ref{fig:group_nov_statistics}{\bf D}). Similar trends are seen for the reference diversity both in terms of its temporal evolution (upper panel of Fig.~S9{\bf A,B}) as well as their temporally averaged values (lower panel). Unlike the entropy or reference diversity, there is a decreasing trend in time for the citation diversity. We observe a 5\% decrease in the measure between those in the top 10\% as compared to the bottom 50\%. Figure~\ref{fig:group_nov_statistics}{\bf B,E} indicates the same trend for Innovation which also decreases in time across all groups, reflecting a saturation in the number of combinations of new terms that are combined by authors as their career progresses. The difference between the top and bottom groups is now around 15\%. Finally, citations to papers experience an initial boost and then decreases in time as seen in Fig.~\ref{fig:group_nov_statistics}{\bf C}, with now much clearer differences between the groups. Indeed, there is a 40\% difference in citations per-paper between the most and least inspired groups as seen in Fig.~\ref{fig:group_nov_statistics}{\bf F}. 

In terms of redundancy, in Fig.~S9{\bf C} we plot the cosine similarity (Eq.~\eqref{eq:sim}. As the figure indicates, across all groups there is a decreasing trend in the temporal evolution of the similarity, yet a clear difference exists, whereby papers published by the top 10\% are on average 8\% more similar to each other in terms of content when compared to the bottom 50\%.  Taken together, the results indicate that groups of authors who cite superstar papers often do get a citation boost as compared to other sets of authors. However, their output is modestly more innovative and equally novel as compared to the rest of the corpus. Rather their content is more redundnant than the remaining sets of authors.    
    
 \begin{figure}[t!]
\includegraphics[width=\linewidth]{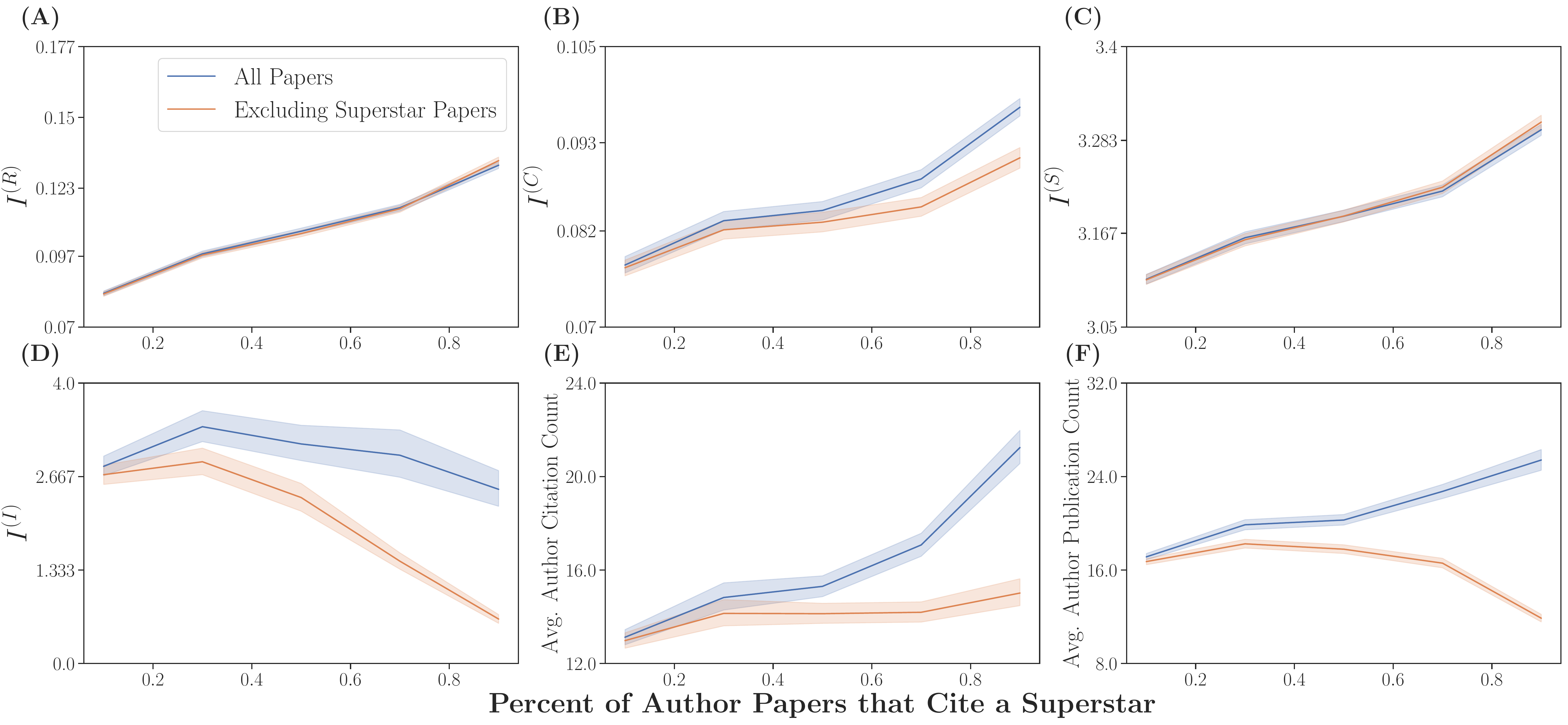}
\caption{{\bf Novelty and Innovation statistics at the individual author-level}. \textbf{A} Reference Diversity, \textbf{B} Citation Diversity, \textbf{C} Shannon Entropy, \textbf{D} Innovation, \textbf{E} Average citation count, \textbf{F} Average publication count.}
\label{fig:individual_nov_stats}
\end{figure}

Next, we dis-aggregate the group-level results and examine the degree of superstar influence at the individual author level. In Fig.~\ref{fig:individual_nov_stats} we plot the averages of the novelty and innovation metrics as well as citations and publication counts across authors as a function of the fraction of their papers that cite superstars. Given that many authors co-publish with superstars, the blue curve indicates the results when including such papers, while the orange curve shows the results excluding these papers. Figure~\ref{fig:individual_nov_stats}{\bf A-C} indicate that as authors cite more superstars they experience an increase in reference and citation diversity as well as the Shannon entropy irrespective of whether one includes their collaboration with superstars. While we see no indications of novelty of content being driven by superstar-influence at the group-level, at the individual level the benefits are clear. On the other hand, when looking at Innovation (Fig.~\ref{fig:individual_nov_stats}{\bf D}), the trend is either flat when including all papers, and decreasing when co-authored publications are excluded. Indeed, it appears that the more authors cite superstars, the \emph{less innovative} their own publications become (i.e those not co-authored with a superstar). The benefit of collaborating with a superstar becomes even more apparent when looking at citations (Fig.~\ref{fig:individual_nov_stats}{\bf E}) and number of publications (Fig.~\ref{fig:individual_nov_stats} {\bf F}). For the former when including collaborations there is a dramatic benefit in terms of garnered citations (approximately 67\% more citations on average) that drops considerably when excluding collaborations. Indeed, the citation-benefit appears to be driven primarily by being collaborators of superstars who by definition have the largest number of citations to their papers. The same appears to be the case for the latter, with the number of publications increasing when including collaborations, and decreasing when excluded.

    \subsection{Early Collaborators and Early Innovators}
    

\begin{figure}[t!]
\includegraphics[width=.95\linewidth]{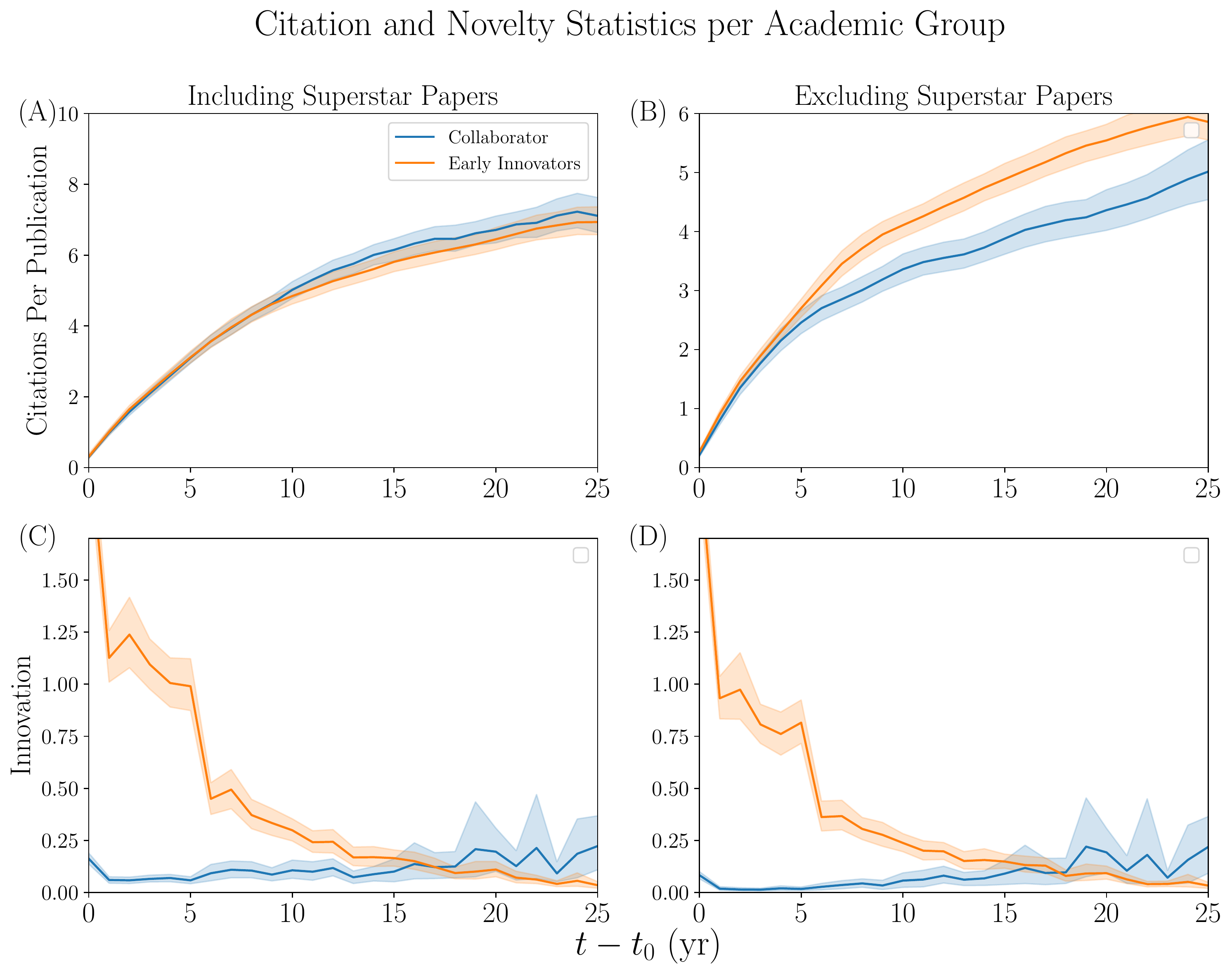}
\centering
\caption{{\bf Citations and Innovation for frequent collaborators and early innovators} \textbf{A} Citations per paper when including superstar papers, \textbf{B} The same when excluding superstar papers. \textbf{C} Temporal evolution of Innovation. \textbf{D} The same when excluding superstar papers. The horizontal axis $t-t_0$ indicates the time elapsed from the $t_0$ the time of first publication for authors in either group.}
\label{fig:early_collab_inov_stats}
\end{figure} 

The results thus far provide evidence for academics inspired by superstars producing output with diverse content and that receives visibility via citations, while not necessarily being innovative in the sense of tying together new concepts. On the other hand, there is also evidence that these features are significantly boosted by direct collaboration with superstars, and when left to their own devices their publication output, novelty and innovation is lower than the rest of the corpus. Indeed, it begs the question whether superstars foster independent individual success, or rather inhibits it? For instance, as shown, at the aggregate level, the group of authors that cite superstars the most often tend to publish on mostly the same topics. 

To further probe this we restrict our analysis to early-career scientists. Given that findings from prior studies have shown that collaboration with successful scientists provides a boost for early career researchers~\cite{top-scientist-collab-success}, and that early success generates a cumulative advantage of long-term career success~\cite{Matthew-Effect}, we define \emph{early collaborators} as those authors who collaborate with superstars in at least half of their papers in the first five years of their career. As a point of comparison, we define another set of authors  who do not collaborate with, or cite superstar papers, but are in the top 10\% of the corpus in terms of Innovation as measured by their first five years of publications. We term these authors \emph{early innovators}. We use innovation as a metric, given that this is the measure by which superstars outperform other academics the most (Fig.~\ref{fig:author_ss_nov_stats}{\bf D}) and therefore might serve as a robust indicator of academic potential.

For academics in each group we track the temporal evolution of the citations per-paper, the number of publications, as well as the Innovation, measured from the date of first publication $t_0$ for authors in either group.  Early collaborators get more citations per paper (Fig.~\ref{fig:early_collab_inov_stats}{\bf A}) and publish more than early innovators (Fig.~S10{\bf A}) particularly within the first ten years of their career. However, when one removes superstar publications, the trend reverses where now early innovators publish more (Fig.~S10{\bf B}) and garner a comparable rate of citations as the other group (Fig.~\ref{fig:early_collab_inov_stats}{\bf B }). Additionally the early innovators maintain a higher degree of Innovation throughout their careers as compared to early collaborators (Fig.~\ref{fig:early_collab_inov_stats}{\bf C, D}) with or without including collaborations to superstars. Thus the evidence suggests that while early career scientists indeed get a boost from collaborating with superstars, their own academic output is less innovative and equally visible in terms of citations, as compared to other early career scientists who produce innovative output without the benefit of such collaborations.

\section{Conclusion and Discussion}
    
    In the exponentially growing knowledge-base of academia in which visibility and funding are increasingly being biased towards top academics and institutions, we examine the influence that superstar academics have on the community as a whole and in terms of novelty and career success.
    Superstars provide an irreplaceable source of novel ideas and contributions at rates that exceed those of other academics in the corpus; our metrics support that their accolades are well deserved and should be rewarded as such.
    We find superstars are highly novel and inspire a higher diversity of concepts among their followers and collaborators. However they do inhibit innovation potential. Those academics most inspired by a superstar are individually themselves more diverse in their papers, but at the group level add little intrinsic novelty than groups more weakly inspired by the superstar, even though they achieve higher citations.
    
    Additionally, we find indications of a strong Matthew Effect whereby academics who cite a superstar highly receive higher citations when collaborating with the superstar than without, despite higher gains in concept diversity than academic counterparts. Though collaboration with successful academics can stimulate a successful career path, we find these collaborations can stifle innovation and may not provide the best indicator of long-term independent career success.
    
    Collaboration is a requirement to tackle increasingly difficult interdisciplinary problems. Superstars are well-positioned to foster interdisciplinary research efforts by supporting early-career researchers. Although the latter receive a citation boost when collaborate with a superstar, this does not imply that they are developing more novel work than their colleagues who are less connected to top academics. In fact, our results indicate that those closest to a superstar show the lowest innovation potential. This is slightly surprising given that the literature have shown junior researchers that collaborate with superstars are more likely to publish in high quality journals and have increased chances of engaging in high quality research with other top scientists. On balance, however, we find that this does not stimulate long term independent career success. This could be an indication of individuals getting lost in the wake of a superstar, meaning these researchers ``bandwagon'' off the ideas and visibility of their respective superstars and iterate on the superstar's work. Although there is value in iterating upon already developed research questions, this may not foster innovative work and stimulate individual careers. Indeed, very recently it has been shown that there is a decline in disruptive ideas in both scientific publications and patents~\cite{Park_2023}. The authors attribute this to an ever increasing reliance on a narrower set of extant scientific knowledge on which to build ideas, a finding very much in line with our observation that followers of superstars produce redundant and less innovative content as a group.

    The observed effects could be a consequence of superstars' strong hold over their respective fields. It's been shown that paradigm shifts in thinking occur after the sudden deaths of superstars. Collaborators of superstars suffer a drop in publication rate after their superstar death, and the field may experience a surge of contributions by outsiders who are disproportionately likely to be highly-cited \cite{Azoulay-funeral}. One can infer that collaborators of superstars are successful because they are collaborating with superstars. Care should be taken when considering these proteges themselves for matters of funding and academic hiring. If the goal is to foster highly novel work, elements outside of prestige and social connection, such as efficacy, equity, and innovation, should be considered.

Our findings are not limited solely to early innovators, collaborators, and inspirees. Though we provide early innovators as an example, many other groups \cite{mining:bing2013} can be isolated and studied in the way we have done here to identify promising academics based on early signatures of novelty or a range of social parameters. We outlined multiple different definitions of novelty in the introduction which we have not further developed in this study. Implementing the different definitions and distinguishing different types of novelty can elucidate what types of novelty are stifled or enhanced by different social configurations.
 
   A subject that we have not probed but is directly relevant to our discussion is the matter of funding. In recent times, funding has increasingly become more biased towards top institutions \cite{funding-bias-canada}, with 90$\%$ of NSF funding in 2018 going to 22$\%$ of funded institutions, serving 43$\%$ of all institutions and 34$\%$ of underrepresented minorities \cite{APS-Report}. This is coupled with a history of funding disparities with respect to race and underrepresented communities \cite{funding-inequalities,racial-disparity-NSF,racial-disparity}. Additionally, underrepresented groups produce novel works at higher rates yet are taken up by other scholars at lower rates than novel contributions by gender and racial majorities \cite{diversity-innovation-paradox}. Equitable funding programs have been shown to enhance research infrastructure, investigator capabilities, and intra- and inter-university collaborations at less prominent institutions \cite{EPSCoR}. As we have shown, those that are least influenced by superstars innovate the most and consequently have higher citation rates. Coupling these results with added attention to equitable funding practices \cite{funding:bollen2014} we believe will reduce the growing inequality in academia and stimulate novel and innovative research.    
    
Finally, we note that our investigation necessarily comes with limitations. Given our sole focus on the APS body of literature, one should be careful to extrapolate this to other academic disciplines. This is also an incomplete subset of the entire journal, so a full corpus with an entire citation network would give a more accurate picture.

   \bibliographystyle{naturemag}
\bibliography{sample}

\begin{thebibliography}{10}
\expandafter\ifx\csname url\endcsname\relax
  \def\url#1{\texttt{#1}}\fi
\expandafter\ifx\csname urlprefix\endcsname\relax\def\urlprefix{URL }\fi
\providecommand{\bibinfo}[2]{#2}
\providecommand{\eprint}[2][]{\url{#2}}

\bibitem{Cliff_2021}
\bibinfo{author}{Cliff, H.}
\newblock \emph{\bibinfo{title}{How to make an Apple Pie From Scratch In Search
  of the Recipe for our Universe}} (\bibinfo{publisher}{Picador},
  \bibinfo{address}{London}, \bibinfo{year}{2021}).

\bibitem{McAndrew:2014yi}
\bibinfo{author}{McAndrew, S.} \& \bibinfo{author}{Everett, M.}
\newblock \bibinfo{title}{Music as collective invention: A social network
  analysis of composers}.
\newblock \emph{\bibinfo{journal}{Cultural Sociology}}
  \textbf{\bibinfo{volume}{9}}, \bibinfo{pages}{56--80} (\bibinfo{year}{2014}).
\newblock \urlprefix\url{https://doi.org/10.1177/1749975514542486}.

\bibitem{Muller:2019ny}
\bibinfo{author}{Muller, E.} \& \bibinfo{author}{Peres, R.}
\newblock \bibinfo{title}{The effect of social networks structure on innovation
  performance: A review and directions for research}.
\newblock \emph{\bibinfo{journal}{International Journal of Research in
  Marketing}} \textbf{\bibinfo{volume}{36}}, \bibinfo{pages}{3--19}
  (\bibinfo{year}{2019}).
\newblock
  \urlprefix\url{https://www.sciencedirect.com/science/article/pii/S0167811618300284}.

\bibitem{Hazarie:2020zb}
\bibinfo{author}{Hazarie, S.}, \bibinfo{author}{Barbosa, H.},
  \bibinfo{author}{Frank, A.}, \bibinfo{author}{Menezes, R.} \&
  \bibinfo{author}{Ghoshal, G.}
\newblock \bibinfo{title}{Uncovering the differences and similarities between
  physical and virtual mobility}.
\newblock \emph{\bibinfo{journal}{Journal of The Royal Society Interface}}
  \textbf{\bibinfo{volume}{17}}, \bibinfo{pages}{20200250}
  (\bibinfo{year}{2020}).
\newblock \urlprefix\url{https://doi.org/10.1098/rsif.2020.0250}.

\bibitem{Chen:2022tt}
\bibinfo{author}{Chen, Z.} \emph{et~al.}
\newblock \bibinfo{title}{Contrasting social and non-social sources of
  predictability in human mobility}.
\newblock \emph{\bibinfo{journal}{Nature Communications}}
  \textbf{\bibinfo{volume}{13}}, \bibinfo{pages}{1922} (\bibinfo{year}{2022}).
\newblock \urlprefix\url{https://doi.org/10.1038/s41467-022-29592-y}.

\bibitem{Rodriguez_2016}
\bibinfo{author}{Nathaniel~Rodriguez, Y.-Y.~A., Johan~Bollen}.
\newblock \bibinfo{title}{Collective dynamics of belief evolution under
  cognitive coherence and social conformity}.
\newblock \emph{\bibinfo{journal}{{PLoS ONE}}} \textbf{\bibinfo{volume}{11}},
  \bibinfo{pages}{e0165910} (\bibinfo{year}{2016}).

\bibitem{Holme:2006bf}
\bibinfo{author}{Holme, P.} \& \bibinfo{author}{Ghoshal, G.}
\newblock \bibinfo{title}{Dynamics of networking agents competing for high
  centrality and low degree}.
\newblock \emph{\bibinfo{journal}{Physical Review Letters}}
  \textbf{\bibinfo{volume}{96}}, \bibinfo{pages}{098701--}
  (\bibinfo{year}{2006}).
\newblock
  \urlprefix\url{https://link.aps.org/doi/10.1103/PhysRevLett.96.098701}.

\bibitem{Ghoshal:2007la}
\bibinfo{author}{Ghoshal, G.} \& \bibinfo{author}{Newman, M. E.~J.}
\newblock \bibinfo{title}{Growing distributed networks with arbitrary degree
  distributions}.
\newblock \emph{\bibinfo{journal}{The European Physical Journal B}}
  \textbf{\bibinfo{volume}{58}}, \bibinfo{pages}{175--184}
  (\bibinfo{year}{2007}).
\newblock \urlprefix\url{https://doi.org/10.1140/epjb/e2007-00208-2}.

\bibitem{Recuero:2019jp}
\bibinfo{author}{Recuero, R.}, \bibinfo{author}{Zago, G.} \&
  \bibinfo{author}{Soares, F.}
\newblock \bibinfo{title}{Using social network analysis and social capital to
  identify user roles on polarized political conversations on twitter}.
\newblock \emph{\bibinfo{journal}{Social Media + Society}}
  \textbf{\bibinfo{volume}{5}}, \bibinfo{pages}{2056305119848745}
  (\bibinfo{year}{2019}).
\newblock \urlprefix\url{https://doi.org/10.1177/2056305119848745}.

\bibitem{Dubois:2014ch}
\bibinfo{author}{Dubois, E.} \& \bibinfo{author}{Gaffney, D.}
\newblock \bibinfo{title}{The multiple facets of influence: Identifying
  political influentials and opinion leaders on twitter}.
\newblock \emph{\bibinfo{journal}{American Behavioral Scientist}}
  \textbf{\bibinfo{volume}{58}}, \bibinfo{pages}{1260--1277}
  (\bibinfo{year}{2014}).
\newblock \urlprefix\url{https://doi.org/10.1177/0002764214527088}.

\bibitem{RADICCHI2017704}
\bibinfo{author}{Radicchi, F.}, \bibinfo{author}{Weissman, A.} \&
  \bibinfo{author}{Bollen, J.}
\newblock \bibinfo{title}{Quantifying perceived impact of scientific
  publications}.
\newblock \emph{\bibinfo{journal}{Journal of Informetrics}}
  \textbf{\bibinfo{volume}{11}}, \bibinfo{pages}{704--712}
  (\bibinfo{year}{2017}).
\newblock
  \urlprefix\url{https://www.sciencedirect.com/science/article/pii/S1751157717300846}.

\bibitem{hindex:hirsch2005}
\bibinfo{author}{Hirsch, J.~E.}
\newblock \bibinfo{title}{An index to quantify an individual's scientific
  research output}.
\newblock \emph{\bibinfo{journal}{Proceedings of the National Academy of
  Sciences}} \textbf{\bibinfo{volume}{102}}, \bibinfo{pages}{16569--16572}
  (\bibinfo{year}{2005}).

\bibitem{global-citation-inequality}
\bibinfo{author}{Nielsen, M.~W.} \& \bibinfo{author}{Andersen, J.~P.}
\newblock \bibinfo{title}{Global citation inequality is on the rise}.
\newblock \emph{\bibinfo{journal}{Proceedings of the National Academy of
  Sciences}} \textbf{\bibinfo{volume}{118}}, \bibinfo{pages}{e2012208118}
  (\bibinfo{year}{2021}).

\bibitem{Matthew-Effect}
\bibinfo{author}{Merton, R.~K.}
\newblock \bibinfo{title}{The matthew effect in science}.
\newblock \emph{\bibinfo{journal}{Science}} \textbf{\bibinfo{volume}{159}},
  \bibinfo{pages}{56--63} (\bibinfo{year}{1968}).

\bibitem{encyclopedia-creativity}
\bibinfo{author}{Runco, M.} \& \bibinfo{author}{Pritzker, S.}
\newblock \emph{\bibinfo{title}{Encyclopedia of Creativity}}.
\newblock Encyclopedia of Creativity (\bibinfo{publisher}{Elsevier Science},
  \bibinfo{year}{2011}).

\bibitem{top-scientist-collab-success}
\bibinfo{author}{Li, W.}, \bibinfo{author}{Aste, T.},
  \bibinfo{author}{Caccioli, F.} \& \bibinfo{author}{Livan, G.}
\newblock \bibinfo{title}{Early coauthorship with top scientists predicts
  success in academic careers}.
\newblock \emph{\bibinfo{journal}{Nature Communications}}
  \textbf{\bibinfo{volume}{10}}, \bibinfo{pages}{5170} (\bibinfo{year}{2019}).

\bibitem{chaperone-effect}
\bibinfo{author}{Sekara, V.} \emph{et~al.}
\newblock \bibinfo{title}{The chaperone effect in scientific publishing}.
\newblock \emph{\bibinfo{journal}{Proceedings of the National Academy of
  Sciences}} \textbf{\bibinfo{volume}{115}}, \bibinfo{pages}{12603--12607}
  (\bibinfo{year}{2018}).

\bibitem{coauthorship-top-ordinary}
\bibinfo{author}{Xie, Q.}, \bibinfo{author}{Zhang, X.}, \bibinfo{author}{Kim,
  G.} \& \bibinfo{author}{Song, M.}
\newblock \bibinfo{title}{Exploring the influence of coauthorship with top
  scientists on researchers' affiliation, research topic, productivity, and
  impact}.
\newblock \emph{\bibinfo{journal}{Journal of Informetrics}}
  \textbf{\bibinfo{volume}{16}}, \bibinfo{pages}{101314}
  (\bibinfo{year}{2022}).
\newblock
  \urlprefix\url{https://www.sciencedirect.com/science/article/pii/S1751157722000669}.

\bibitem{Abrahamson-innovation-diffusion}
\bibinfo{author}{Abrahamson, E.} \& \bibinfo{author}{Rosenkopf, L.}
\newblock \bibinfo{title}{Social network effects on the extent of innovation
  diffusion: A computer simulation}.
\newblock \emph{\bibinfo{journal}{Organization Science}}
  \textbf{\bibinfo{volume}{8}}, \bibinfo{pages}{289--309}
  (\bibinfo{year}{1997}).
\newblock \urlprefix\url{http://www.jstor.org/stable/2635149}.

\bibitem{Superstar-Extinction}
\bibinfo{author}{Azoulay, P.}, \bibinfo{author}{Graff~Zivin, J.~S.} \&
  \bibinfo{author}{Wang, J.}
\newblock \bibinfo{title}{{Superstar Extinction}}.
\newblock \emph{\bibinfo{journal}{The Quarterly Journal of Economics}}
  \textbf{\bibinfo{volume}{125}}, \bibinfo{pages}{549--589}
  (\bibinfo{year}{2010}).
\newblock \urlprefix\url{https://doi.org/10.1162/qjec.2010.125.2.549}.
\newblock
  \eprint{https://academic.oup.com/qje/article-pdf/125/2/549/5319678/125-2-549.pdf}.

\bibitem{Clauset:eu}
\bibinfo{author}{Clauset, A.}, \bibinfo{author}{Arbesman, S.} \&
  \bibinfo{author}{Larremore, D.~B.}
\newblock \bibinfo{title}{Systematic inequality and hierarchy in faculty hiring
  networks}.
\newblock \emph{\bibinfo{journal}{Science Advances}}
  \textbf{\bibinfo{volume}{1}}, \bibinfo{pages}{e1400005}
  (\bibinfo{year}{2015}).
\newblock \urlprefix\url{https://doi.org/10.1126/sciadv.1400005}.

\bibitem{Janosov:2020ri}
\bibinfo{author}{Janosov, M.}, \bibinfo{author}{Battiston, F.} \&
  \bibinfo{author}{Sinatra, R.}
\newblock \bibinfo{title}{Success and luck in creative careers}.
\newblock \emph{\bibinfo{journal}{EPJ Data Science}}
  \textbf{\bibinfo{volume}{9}}, \bibinfo{pages}{9} (\bibinfo{year}{2020}).
\newblock \urlprefix\url{https://doi.org/10.1140/epjds/s13688-020-00227-w}.

\bibitem{Matthew-Effect-Science-Funding}
\bibinfo{author}{Bol, T.}, \bibinfo{author}{de~Vaan, M.} \&
  \bibinfo{author}{van~de Rijt, A.}
\newblock \bibinfo{title}{The matthew effect in science funding}.
\newblock \emph{\bibinfo{journal}{Proceedings of the National Academy of
  Sciences}} \textbf{\bibinfo{volume}{115}}, \bibinfo{pages}{4887--4890}
  (\bibinfo{year}{2018}).
\newblock \urlprefix\url{https://www.pnas.org/doi/abs/10.1073/pnas.1719557115}.
\newblock \eprint{https://www.pnas.org/doi/pdf/10.1073/pnas.1719557115}.

\bibitem{Matthew-Effect-Model}
\bibinfo{author}{Petersen, A.~M.}, \bibinfo{author}{Jung, W.-S.},
  \bibinfo{author}{Yang, J.-S.} \& \bibinfo{author}{Stanley, H.~E.}
\newblock \bibinfo{title}{Quantitative and empirical demonstration of the
  matthew effect in a study of career longevity}.
\newblock \emph{\bibinfo{journal}{Proceedings of the National Academy of
  Sciences}} \textbf{\bibinfo{volume}{108}}, \bibinfo{pages}{18--23}
  (\bibinfo{year}{2011}).
\newblock \urlprefix\url{https://www.pnas.org/doi/abs/10.1073/pnas.1016733108}.
\newblock \eprint{https://www.pnas.org/doi/pdf/10.1073/pnas.1016733108}.

\bibitem{exploitation-exploration-Lazer}
\bibinfo{author}{Lazer, D.} \& \bibinfo{author}{Friedman, A.}
\newblock \bibinfo{title}{The network structure of exploration and
  exploitation}.
\newblock \emph{\bibinfo{journal}{Administrative Science Quarterly}}
  \textbf{\bibinfo{volume}{52}}, \bibinfo{pages}{667 -- 694}
  (\bibinfo{year}{2007}).

\bibitem{Rodan-Heterogeneity-Innovation}
\bibinfo{author}{Rodan, S.} \& \bibinfo{author}{Galunic, C.}
\newblock \bibinfo{title}{More than network structure: How knowledge
  heterogeneity influences managerial performance and innovativeness}.
\newblock \emph{\bibinfo{journal}{Strategic Management Journal}}
  \textbf{\bibinfo{volume}{25}}, \bibinfo{pages}{541--562}
  (\bibinfo{year}{2004}).
\newblock \urlprefix\url{http://www.jstor.org/stable/20142143}.

\bibitem{Chang-knowledge-diffusion-imitation-innovation}
\bibinfo{author}{Chang, M.} \& \bibinfo{author}{Joseph E.~Harrington, J.}
\newblock \bibinfo{title}{Discovery and diffusion of knowledge in an endogenous
  social network}.
\newblock \emph{\bibinfo{journal}{American Journal of Sociology}}
  \textbf{\bibinfo{volume}{110}}, \bibinfo{pages}{937--976}
  (\bibinfo{year}{2005}).
\newblock \urlprefix\url{http://www.jstor.org/stable/10.1086/426555}.

\bibitem{Novelty-Recognition-MatthewEffect}
\bibinfo{author}{Trapido, D.}
\newblock \bibinfo{title}{How novelty in knowledge earns recognition: The role
  of consistent identities}.
\newblock \emph{\bibinfo{journal}{Research Policy}}
  \textbf{\bibinfo{volume}{44}}, \bibinfo{pages}{1488--1500}
  (\bibinfo{year}{2015}).
\newblock
  \urlprefix\url{https://www.sciencedirect.com/science/article/pii/S0048733315000839}.

\bibitem{flat:evans2022}
\bibinfo{author}{Xu, F.} \& \bibinfo{author}{Evans, J.}
\newblock \bibinfo{title}{Flat teams drive scientific innovation}.
\newblock \emph{\bibinfo{journal}{Proceedings of the National Academy of
  Sciences}} \textbf{\bibinfo{volume}{119}} (\bibinfo{year}{2022}).

\bibitem{h-index-justification}
\bibinfo{author}{Hirsch, J.~E.}
\newblock \bibinfo{title}{Does the h-index have predictive power?}
\newblock \emph{\bibinfo{journal}{Proceedings of the National Academy of
  Sciences}} \textbf{\bibinfo{volume}{104}}, \bibinfo{pages}{19193--19198}
  (\bibinfo{year}{2007}).

\bibitem{h-index}
\bibinfo{author}{Hirsch, J.~E.}
\newblock \bibinfo{title}{An index to quantify an individual's scientific
  research output}.
\newblock \emph{\bibinfo{journal}{Proceedings of the National Academy of
  Sciences}} \textbf{\bibinfo{volume}{102}}, \bibinfo{pages}{16569--16572}
  (\bibinfo{year}{2005}).
\newblock \urlprefix\url{https://www.pnas.org/doi/abs/10.1073/pnas.0507655102}.
\newblock \eprint{https://www.pnas.org/doi/pdf/10.1073/pnas.0507655102}.

\bibitem{APS-dataset}
\bibinfo{author}{{American Physical Society}}.
\newblock \bibinfo{howpublished}{\url{https://journals.aps.org/datasets}}.

\bibitem{beautiful-soup}
\bibinfo{author}{Richardson, L.}
\newblock
  \bibinfo{howpublished}{\url{https://sethc23.github.io/wiki/Python/Beautiful_Soup_Documentation.pdf}}.

\bibitem{author-disambiguation}
\bibinfo{author}{Caron, E.} \& \bibinfo{author}{{van Eck}, N.-J.}
\newblock \bibinfo{title}{Large scale author name disambiguation using
  rule-based scoring and clustering}.
\newblock In \bibinfo{editor}{Noyons, E.} (ed.)
  \emph{\bibinfo{booktitle}{Proceedings of the Science and Technology
  Indicators Conference 2014}}, \bibinfo{pages}{79--86}
  (\bibinfo{publisher}{Universiteit Leiden}, \bibinfo{year}{2014}).
\newblock \urlprefix\url{http://sti2014.cwts.nl}.
\newblock \bibinfo{note}{International conference on science and technology
  indicators, STI 2014 ; Conference date: 03-09-2014 Through 05-09-2014}.

\bibitem{PLDA}
\bibinfo{author}{El-Kishky, A.}, \bibinfo{author}{Song, Y.},
  \bibinfo{author}{Wang, C.}, \bibinfo{author}{Voss, C.~R.} \&
  \bibinfo{author}{Han, J.}
\newblock \bibinfo{title}{Scalable topical phrase mining from text corpora}.
\newblock \emph{\bibinfo{journal}{Proc. VLDB Endow.}}
  \textbf{\bibinfo{volume}{8}}, \bibinfo{pages}{305--316}
  (\bibinfo{year}{2014}).
\newblock \urlprefix\url{https://doi.org/10.14778/2735508.2735519}.

\bibitem{Lee:2022tf}
\bibinfo{author}{Lee, S.~Y.}
\newblock \bibinfo{title}{Gibbs sampler and coordinate ascent variational
  inference: A set-theoretical review}.
\newblock \emph{\bibinfo{journal}{Communications in Statistics - Theory and
  Methods}} \textbf{\bibinfo{volume}{51}}, \bibinfo{pages}{1549--1568}
  (\bibinfo{year}{2022}).
\newblock \urlprefix\url{https://doi.org/10.1080/03610926.2021.1921214}.

\bibitem{UMASS-Coherence}
\bibinfo{author}{Mimno, D.}, \bibinfo{author}{Wallach, H.},
  \bibinfo{author}{Talley, E.}, \bibinfo{author}{Leenders, M.} \&
  \bibinfo{author}{McCallum, A.}
\newblock \bibinfo{title}{Optimizing semantic coherence in topic models}.
\newblock In \emph{\bibinfo{booktitle}{Proceedings of the 2011 Conference on
  Empirical Methods in Natural Language Processing}}, \bibinfo{pages}{262--272}
  (\bibinfo{publisher}{Association for Computational Linguistics},
  \bibinfo{address}{Edinburgh, Scotland, UK.}, \bibinfo{year}{2011}).
\newblock \urlprefix\url{https://aclanthology.org/D11-1024}.

\bibitem{Novelty-Detection-Review}
\bibinfo{author}{Ouafae, B.}, \bibinfo{author}{Oumaima, L.},
  \bibinfo{author}{Mariam, R.} \& \bibinfo{author}{Abdelouahid, L.}
\newblock \bibinfo{title}{Novelty detection review state of art and discussion
  of new innovations in the main application domains}.
\newblock In \emph{\bibinfo{booktitle}{2020 1st International Conference on
  Innovative Research in Applied Science, Engineering and Technology
  (IRASET)}}, \bibinfo{pages}{1--7} (\bibinfo{year}{2020}).

\bibitem{Novelty-Detecion-TREC}
\bibinfo{author}{Soboroff, I.} \& \bibinfo{author}{Harman, D.}
\newblock \bibinfo{title}{Overview of the {TREC} 2003 novelty track}.
\newblock In \bibinfo{editor}{Voorhees, E.~M.} \& \bibinfo{editor}{Buckland,
  L.~P.} (eds.) \emph{\bibinfo{booktitle}{Proceedings of The Twelfth Text
  REtrieval Conference, {TREC} 2003, Gaithersburg, Maryland, USA, November
  18-21, 2003}}, vol. \bibinfo{volume}{500-255} of
  \emph{\bibinfo{series}{{NIST} Special Publication}}, \bibinfo{pages}{38--53}
  (\bibinfo{publisher}{National Institute of Standards and Technology
  {(NIST)}}, \bibinfo{year}{2003}).
\newblock
  \urlprefix\url{http://trec.nist.gov/pubs/trec12/papers/NOVELTY.OVERVIEW.pdf}.

\bibitem{Ghoshal-Novelty-Detection}
\bibinfo{author}{Ghosal, T.}, \bibinfo{author}{Saikh, T.},
  \bibinfo{author}{Biswas, T.}, \bibinfo{author}{Ekbal, A.} \&
  \bibinfo{author}{Bhattacharyya, P.}
\newblock \bibinfo{title}{{Novelty Detection: A Perspective from Natural
  Language Processing}}.
\newblock \emph{\bibinfo{journal}{Computational Linguistics}}
  \textbf{\bibinfo{volume}{48}}, \bibinfo{pages}{77--117}
  (\bibinfo{year}{2022}).
\newblock \urlprefix\url{https://doi.org/10.1162/coli\_a\_00429}.
\newblock
  \eprint{https://direct.mit.edu/coli/article-pdf/48/1/77/2006641/coli\_a\_00429.pdf}.

\bibitem{Uzzi-Atypical-Recomb}
\bibinfo{author}{Uzzi, B.}, \bibinfo{author}{Mukherjee, S.},
  \bibinfo{author}{Stringer, M.} \& \bibinfo{author}{Jones, B.}
\newblock \bibinfo{title}{Atypical combinations and scientific impact}.
\newblock \emph{\bibinfo{journal}{Science}} \textbf{\bibinfo{volume}{342}},
  \bibinfo{pages}{468--472} (\bibinfo{year}{2013}).
\newblock
  \urlprefix\url{https://www.science.org/doi/abs/10.1126/science.1240474}.
\newblock \eprint{https://www.science.org/doi/pdf/10.1126/science.1240474}.

\bibitem{Schumpeter:1982}
\bibinfo{author}{Schumpeter, J.~A.}
\newblock \emph{\bibinfo{title}{The theory of economic development: An inquiry
  into profits, capital, credit, interest, and the business cycle (Theorie der
  wirtschaftlichen Entwicklung)}} (\bibinfo{publisher}{Transaction},
  \bibinfo{address}{Edison, NJ}, \bibinfo{year}{1934}).
\newblock \bibinfo{note}{Translated by Redvers Opie}.

\bibitem{Information-Theory-Book}
\bibinfo{author}{Cover, T.} \& \bibinfo{author}{Thomas, J.~A.}
\newblock \emph{\bibinfo{title}{Elements of Information Theory}}.
\newblock Wiley Series in Telecommunications and Signal Processing
  (\bibinfo{publisher}{Wiley-Interscience}, \bibinfo{address}{New York, New
  York, USA}, \bibinfo{year}{2006}).

\bibitem{Aral-Dhillon}
\bibinfo{author}{Aral, S.} \& \bibinfo{author}{Dhillon, P.}
\newblock \bibinfo{title}{What (exactly) is novelty in networks? unpacking the
  vision advantages of brokers, bridges, and weak ties.}
\newblock \emph{\bibinfo{journal}{Institute for Operations Research and the
  Management Sciences (INFORMS)}}  (\bibinfo{year}{2021}).
\newblock \urlprefix\url{http://dx.doi.org/10.2139/ssrn.2388254}.
\newblock \eprint{https://ssrn.com/abstract=2388254}.

\bibitem{Kuhn1962}
\bibinfo{author}{Kuhn, T.~S.}
\newblock \emph{\bibinfo{title}{The Structure of Scientific Revolutions}}
  (\bibinfo{publisher}{University of Chicago Press},
  \bibinfo{address}{Chicago}, \bibinfo{year}{1962}).

\bibitem{diversity-innovation-paradox}
\bibinfo{author}{Hofstra, B.} \emph{et~al.}
\newblock \bibinfo{title}{The diversityx2013;innovation paradox in science}.
\newblock \emph{\bibinfo{journal}{Proceedings of the National Academy of
  Sciences}} \textbf{\bibinfo{volume}{117}}, \bibinfo{pages}{9284--9291}
  (\bibinfo{year}{2020}).
\newblock \urlprefix\url{https://www.pnas.org/doi/abs/10.1073/pnas.1915378117}.
\newblock \eprint{https://www.pnas.org/doi/pdf/10.1073/pnas.1915378117}.

\bibitem{Baten-divergent-thinking}
\bibinfo{author}{Baten, R.~A.} \emph{et~al.}
\newblock \bibinfo{title}{Creativity in temporal social networks: how divergent
  thinking is impacted by one's choice of peers}.
\newblock \emph{\bibinfo{journal}{Journal of The Royal Society Interface}}
  \textbf{\bibinfo{volume}{17}}, \bibinfo{pages}{20200667}
  (\bibinfo{year}{2020}).

\bibitem{Baten_2021}
\bibinfo{author}{Baten, R.~A.}, \bibinfo{author}{Aslin, R.~N.},
  \bibinfo{author}{Ghoshal, G.} \& \bibinfo{author}{Hoque, E.}
\newblock \bibinfo{title}{Cues to gender and racial identity reduce creativity
  in diverse social networks}.
\newblock \emph{\bibinfo{journal}{Scientific Reports}}
  \textbf{\bibinfo{volume}{11}}, \bibinfo{pages}{10261} (\bibinfo{year}{2021}).
\newblock \urlprefix\url{https://doi.org/10.1038/s41598-021-89498-5}.

\bibitem{Baten_2022}
\bibinfo{author}{Baten, R.~A.}, \bibinfo{author}{Aslin, R.~N.},
  \bibinfo{author}{Ghoshal, G.} \& \bibinfo{author}{Hoque, M.~E.}
\newblock \bibinfo{title}{Novel idea generation in social networks is optimized
  by exposure to a ``goldilocks'' level of idea-variability}.
\newblock \emph{\bibinfo{journal}{PNAS Nexus}} \textbf{\bibinfo{volume}{1}},
  \bibinfo{pages}{pgac255} (\bibinfo{year}{2022}).

\bibitem{Park_2023}
\bibinfo{author}{Park, M.}, \bibinfo{author}{Leahey, E.} \&
  \bibinfo{author}{Funk, R.~J.}
\newblock \bibinfo{title}{Papers and patents are becoming less disruptive over
  time}.
\newblock \emph{\bibinfo{journal}{Nature}} \textbf{\bibinfo{volume}{613}},
  \bibinfo{pages}{138--144} (\bibinfo{year}{2023}).
\newblock \urlprefix\url{https://doi.org/10.1038/s41586-022-05543-x}.

\bibitem{Azoulay-funeral}
\bibinfo{author}{Azoulay, P.}, \bibinfo{author}{Fons-Rosen, C.} \&
  \bibinfo{author}{Graff~Zivin, J.~S.}
\newblock \bibinfo{title}{Does science advance one funeral at a time?}
\newblock \emph{\bibinfo{journal}{American Economic Review}}
  \textbf{\bibinfo{volume}{109}}, \bibinfo{pages}{2889--2920}
  (\bibinfo{year}{2019}).
\newblock
  \urlprefix\url{https://www.aeaweb.org/articles?id=10.1257/aer.20161574}.

\bibitem{mining:bing2013}
\bibinfo{author}{He, B.}, \bibinfo{author}{Ding, Y.}, \bibinfo{author}{Tang,
  J.}, \bibinfo{author}{Reguramalingam, V.} \& \bibinfo{author}{Bollen, J.}
\newblock \bibinfo{title}{{Mining diversity subgraph in multidisciplinary
  scientific collaboration networks: A meso perspective}}.
\newblock \emph{\bibinfo{journal}{Journal of Informetrics}}
  \textbf{\bibinfo{volume}{7}}, \bibinfo{pages}{117--128}
  (\bibinfo{year}{2013}).

\bibitem{funding-bias-canada}
\bibinfo{author}{Murray, D.~L.} \emph{et~al.}
\newblock \bibinfo{title}{Bias in research grant evaluation has dire
  consequences for small universities}.
\newblock \emph{\bibinfo{journal}{PLOS ONE}} \textbf{\bibinfo{volume}{11}},
  \bibinfo{pages}{1--19} (\bibinfo{year}{2016}).
\newblock \urlprefix\url{https://doi.org/10.1371/journal.pone.0155876}.

\bibitem{APS-Report}
\bibinfo{author}{of~Government~Affairs, O.}
\newblock \bibinfo{title}{Building america's stem workforce: Eliminating
  barriers and unlocking advantages}.
\newblock \bibinfo{type}{Tech. Rep.}, \bibinfo{institution}{American Physical
  Society}, \bibinfo{address}{1 Physics Ellipse, College Park, MD 20740-3844}
  (\bibinfo{year}{2021}).

\bibitem{funding-inequalities}
\bibinfo{author}{Woodson, T.} \& \bibinfo{author}{Boutilier, S.}
\newblock \bibinfo{title}{{Impacts for whom? Assessing inequalities in
  NSF-funded broader impacts using the Inclusion-Immediacy Criterion}}.
\newblock \emph{\bibinfo{journal}{Science and Public Policy}}
  \textbf{\bibinfo{volume}{49}}, \bibinfo{pages}{168--178}
  (\bibinfo{year}{2021}).
\newblock \urlprefix\url{https://doi.org/10.1093/scipol/scab072}.
\newblock
  \eprint{https://academic.oup.com/spp/article-pdf/49/2/168/43395599/scab072.pdf}.

\bibitem{racial-disparity-NSF}
\bibinfo{author}{Chen, C.~Y.} \emph{et~al.}
\newblock \bibinfo{title}{Decades of systemic racial disparities in funding
  rates at the national science foundation} (\bibinfo{year}{2022}).
\newblock \urlprefix\url{osf.io/xb57u}.

\bibitem{racial-disparity}
\bibinfo{author}{Ginther, D.} \emph{et~al.}
\newblock \bibinfo{title}{Race, ethnicity, and nih research awards}.
\newblock \emph{\bibinfo{journal}{Science (New York, N.Y.)}}
  \textbf{\bibinfo{volume}{333}}, \bibinfo{pages}{1015--9}
  (\bibinfo{year}{2011}).

\bibitem{EPSCoR}
\bibinfo{author}{Harris, L.~A.}
\newblock \bibinfo{title}{Established program to stimulate competitive research
  (epscor): Background and selected issues}.
\newblock \bibinfo{type}{Tech. Rep.} \bibinfo{number}{R44689},
  \bibinfo{institution}{Congressional Research Service}, \bibinfo{address}{1
  Physics Ellipse, College Park, MD 20740-3844} (\bibinfo{year}{2017}).

\bibitem{funding:bollen2014}
\bibinfo{author}{Bollen, J.}, \bibinfo{author}{Crandall, D.},
  \bibinfo{author}{Junk, D.}, \bibinfo{author}{Ding, Y.} \&
  \bibinfo{author}{B{\"o}rner, K.}
\newblock \bibinfo{title}{From funding agencies to scientific agency}.
\newblock \emph{\bibinfo{journal}{EMBO reports}} \textbf{\bibinfo{volume}{15}},
  \bibinfo{pages}{131--133} (\bibinfo{year}{2014}).
\newblock \urlprefix\url{https://doi.org/10.1002/embr.201338068}.

\end{thebibliography}

\end{document}


\makeatletter
\renewcommand{\maketitle}{\bgroup\setlength{\parindent}{0pt}
	\begin{flushleft}
{\huge Supplementary Information}\\[0.3cm]
		{\large\textbf \@title}
		
		\@author
		
	\end{flushleft}\egroup
}
\makeatother
\title{Creativity and Production in Academic Social Networks}

\newcommand{\floor}[1]{\lfloor #1 \rfloor}

\author{Sean Kelty, Raiyan Abdul Baten, Adiba Proma, Ehsan Hoque, Johann Bollen, Gourab Ghoshal}                         

\date{}

\maketitle

\tableofcontents

\newpage

\section{Data}
\label{sec:dataset}

\subsection{Summary statistics}

\begin{figure}[h!]
\begin{longtable}[c]{|c|c|}
\caption{\textbf{Summary of citation statistics of corpus with and without abstracts}.}
\label{sitab:dataset}\\
\hline
%
%
No. Dois                    & 678,916   \\ \hline
No. Dois (w/ abstracts)                     & 250,628    \\ \hline
No. Authors (after disambiguation)                      & 307,894     \\ \hline
No. Superstars                  & 303      \\ \hline
h-index cutoff for superstars                  & 21      \\ \hline
Avg. h-index                  & 1.74      \\ \hline
Avg. No. References per paper                  & 13.5      \\ \hline
Avg. No. References per paper (w/ abstracts)                  & 5.57      \\ \hline
Avg. No. Citations per paper                  & 14.4      \\ \hline
Avg. No. Citations per paper (w/ abstracts)                & 6.88      \\ \hline

\end{longtable}
\end{figure}

\begin{figure}[h!]
\includegraphics[width=0.8\linewidth]{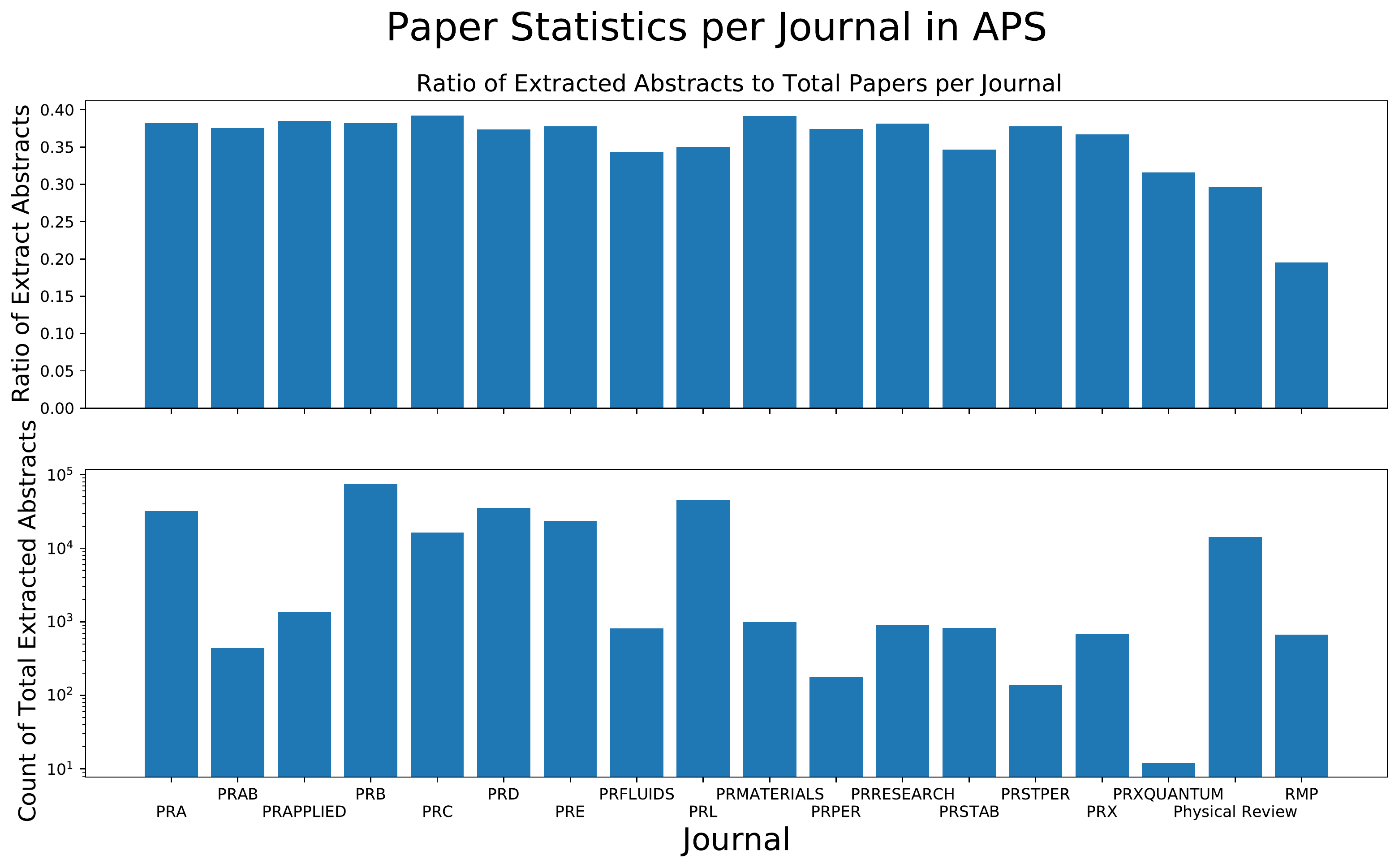}
\caption{Upper panel: Proportion of analyzed papers in across APS journals. Lower panel: Count of all papers in each journal.}
\label{sifig:aps_paper_stats}
\end{figure}

\begin{figure}[h!]
\includegraphics[width=0.8\linewidth]{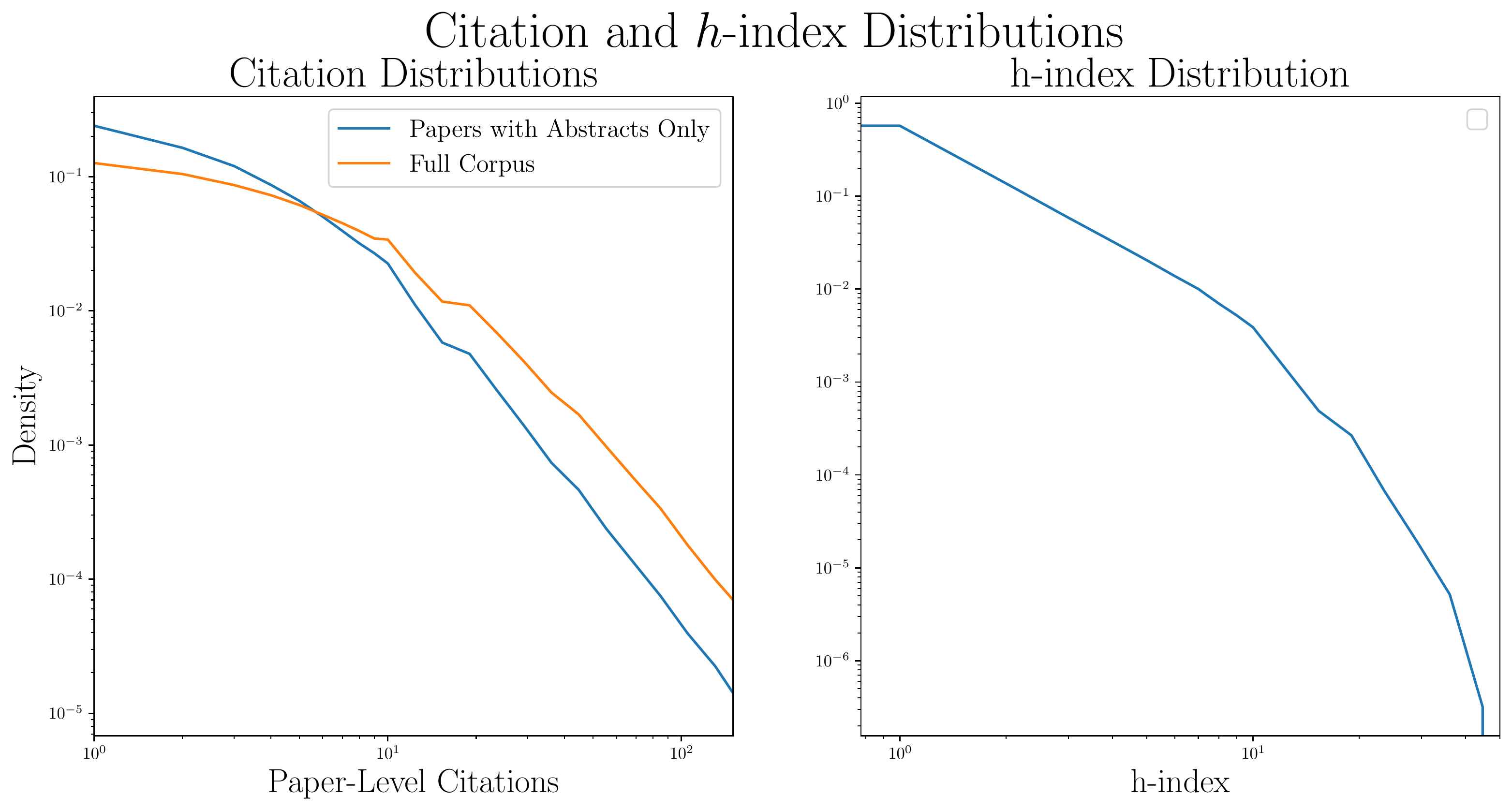}
\caption{Left: Distribution of citations for all papers (orange curve), for papers with extracted abstracts (blue curve). Right: Distribution of the h-index across all authors in the corpus.}
\label{sifig:citation_h_index_distributions}
\end{figure}

\clearpage
\subsection{Author Disambiguation}\label{section:secauthor_disambig}
We use a scoring-based method to disambiguate authors in our dataset.
\begin{enumerate}[noitemsep]
    \item Initials
    \begin{itemize}[noitemsep]
        \item Two Initials : 5
        \item More than 2 initials : 10
        \item Conflicting Initials : -10
    \end{itemize}
    \item First Name
    \begin{itemize}[noitemsep]
        \item General Name : 3
        \item Non-General Name : 6
        
        (A name is considered general if it has been seen more than 1000 times)
    \end{itemize}
    \item Address/Affiliation
    \begin{itemize}[noitemsep]
        \item Country,City : 4
        \item Country,City,Organization : 7
        \item Country,City,Organization,Department : 10
    \end{itemize}
    \item Shared Co-Authors
        \begin{itemize}[noitemsep]
            \item one : 4
            \item two : 7
            \item more than 2 : 10
        \end{itemize}
    \item Source
        \begin{itemize}[noitemsep]
            \item Journal : 6
        \end{itemize}
    \item Self-Citation : 10
    \item Bibliographic Coupling (two works referencing a common third work)
    \begin{itemize}[noitemsep]
        \item one : 2
        \item two : 4
        \item three : 6
        \item four : 8
        \item More than four : 10
    \end{itemize}
    \item Co-citation (Number of times a third work has cited two works)
    \begin{itemize}[noitemsep]
        \item one : 2
        \item two : 3
        \item three : 4
        \item four : 5
        \item More than 4 : 6
    \end{itemize}
\end{enumerate}

\subsection{Coherence of Topic Model}

\begin{figure}[h!]
\includegraphics[width=0.6\linewidth]{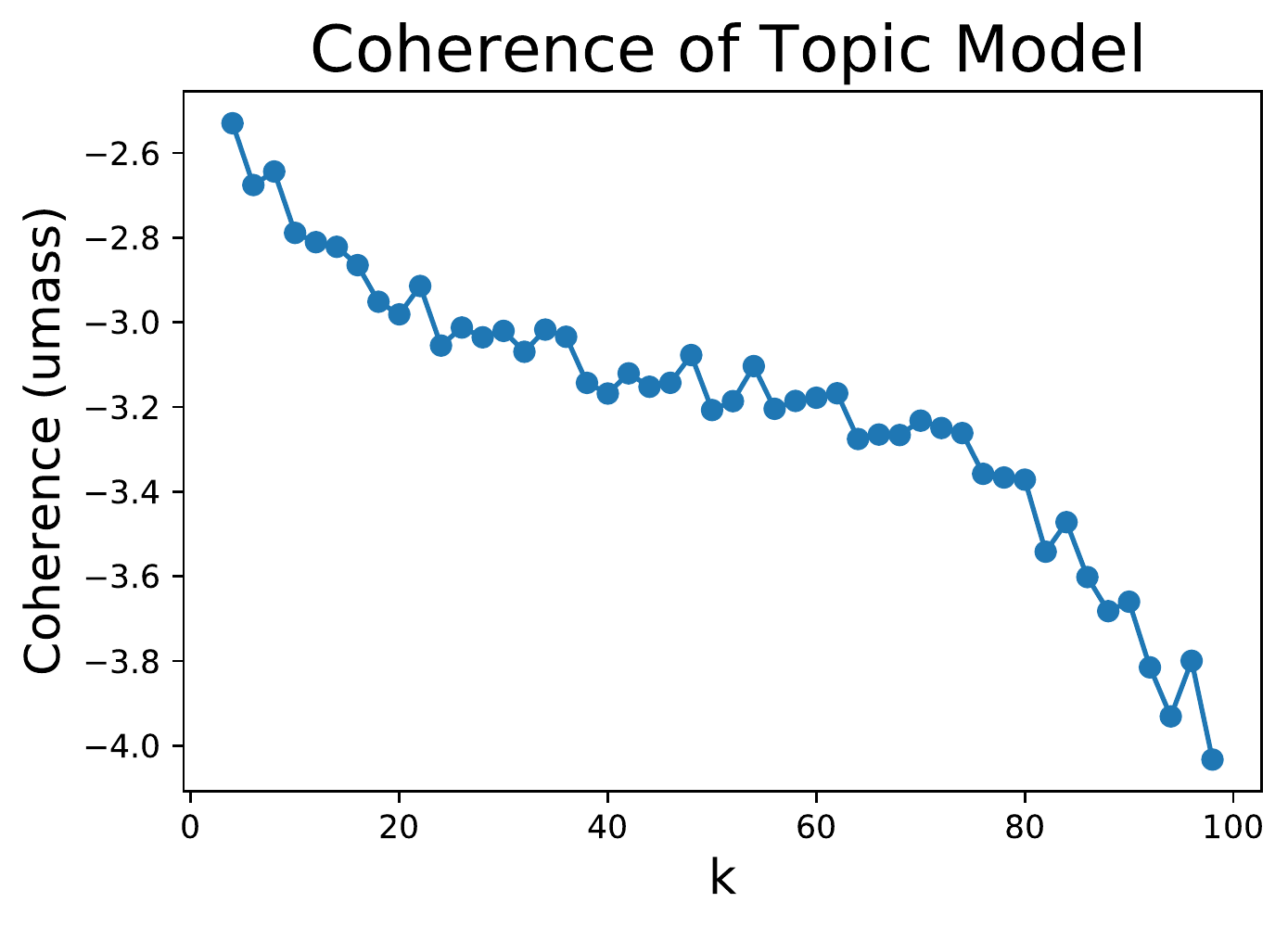}
\caption{Coherence Scores of P-LDA Topic Model}
\label{sifig:coherence_topic_model}
\end{figure}

We apply the UMass coherence measure to determine a stable number of topics for our topic model. This coherence score measures how similar the top words in a topic are to each other. We aim for the highest possible coherence value that stabilizes in a neighborhood of the number of topics $k$. Fig.~\ref{sifig:coherence_topic_model} shows the coherence stablizing at roughly $k = 25$ topics. 
\clearpage

\subsection{Example of Topic Representation}

Words and phrases in the corpus, which will generally be referred to as "terms", are represented by a distribution over latent topics that is the frequency of topic assignments of the term over the entire corpus. Topics are characterized by the frequency of terms associated with the topic. For each topic, all terms are ranked based on their relative topic frequency of their own distribution of the given topic. For example, if a phrase had a topic distribution for $k=3$ topics of [.1,.2,.7], the phrase is representative of topic 3. Terms are pre-processed by removing stop words and stemming words such that conjugated versions of the same word can be represented as the same word.

\begin{longtable}[c]{p{0.1\linewidth} | p{0.9\linewidth}}
\caption{\textbf{Topic Model Summary of most representative terms per topic}.}
\label{sitab:topics}\\
\multicolumn{1}{p{.1\linewidth}|}{Topic Number} & Representative Terms   
\endfirsthead
\hline
\
%
Topic 1      &\footnotesize  crystal film, ultrathin, mtj, stm tip, stack, freestand, high resolution angle, franz, stm, force micrscop   \\ \hline        
Topic 2      & \footnotesize center cubic fcc, temperature addit, measur x, tc cuprat, temperature down k, temperature k k, tc k, tc superconduct, tc superconductor, temperature tc k   \\ \hline   
Topic 3      & \footnotesize spectral line, $\omega$p, raman line, absorpt part, absorpt line, nd3, electroreflect. eliashberg, b1g, endor   \\ \hline   
Topic 4      & \footnotesize axial magnet, spin angular, moment inertia, moment magnet, parallel magnet field, magnet revers, torqu, interlay exchange, spin texture, moriya   \\ \hline   
Topic 5      &  \footnotesize collim, electron eject, ion yield, ion trap, n4, ion produc, ion plasma, damag, wall carbon, electron drift   \\ \hline   
Topic 6      & \footnotesize cauchi, broken time, takahashi, hamilton jacobi, symmetri spontan, tachyon, ward ident, polyakov, loop quantum cosmolog, coulomb guage  \\ \hline   
Topic 7      & \footnotesize excitatori, hub, infect, epidem, volatil, exactli solvabl model, network model, synaps, synapt, integr fire    \\ \hline   
Topic 8      & \footnotesize nonequilibrium phase transit, first order phase transit, j', glass order, thouless transit, glass like, glass former, triangluar lattic, nearest neighbor coupl, nearest neighbor distanc   \\ \hline   
Topic 9      &\footnotesize  magnitude higher, larg part, fourth gener, even though, order qcd, select rule, third, mach zehnder interferomet, even larger, order raman   \\ \hline        
Topic 10      & \footnotesize quasilinear, langevin equat, gilbert equat, equate state eo, sand, attractor, classic chaotic, eulerian, chimera state, euler equat   \\ \hline   
Topic 11      & \footnotesize advanc ligo, mit bag, catalog, model background, dark sector, dark matter, sight, model dark, sky, sno   \\ \hline   
Topic 12      & \footnotesize nest, der waal force, nodal line, helic edg, non fermi, state degeneraci, hove, majorana zero, majorana bound, sdh  \\ \hline   
Topic 13      &  \footnotesize three dimension 3d, basin attract, fuld ferrel, dimension squar, lz, trap bose, bodi effect, bodi forc, hard core boson, fermion atom   \\ \hline   
Topic 14      & \footnotesize highest occupi molecular, muffin tin orbit, gaas1, clathrat, cl2, cl, hexagon boron, interstiti, gell, ci  \\ \hline   
Topic 15      & \footnotesize puls width, optic parametr, sapphir laser, exciton biexciton, optic pump, harmon gener shg, optic puls, inxga1 xa, optic nonlinear, ultrastrong    \\ \hline   
Topic 16      & \footnotesize clauser, horn shimoni holt, simpl analyt express, us deriv, part paper, analyt formula, cb, exact forumla, exact expression, pauli exclus    \\ \hline   
Topic 17      & \footnotesize agre reason, foudn good agreement, recent experiment data, find excel agreement, find good agreement, theoret data, theoret cross, reason agreement experiment, found excel agreement, good agreement experimental result  \\ \hline  
Topic 18      &\footnotesize  qutrit, regist, processor, studi entagle, protocol, markovian dynam, purif, decoy state, qkd, error correct   \\ \hline        
Topic 19      & \footnotesize nucleon nucleon scatter, deep inelast scatter, total angular momentum, inclus cross, transfer cross section, multifragment, multiperipher, depend cross section, forward angle, $\pi$n   \\ \hline   
Topic 20      & \footnotesize full potenti linear augment plane wave, wave born, wannier function, impuls, infield, use path, use mont, within densiti function, jastrow, use harte   \\ \hline   
Topic 21      & \footnotesize avoid walk, nonergod, time $\tau$, time tail, time t2, time t2, dimension diffus, time random, nonexponenti, msd   \\ \hline   
Topic 22      &  \footnotesize even even nuclei, xe136, rich nuclei, gt, v', p0, cf252, $\alpha$ p, $\alpha$ reaction, p1  \\ \hline   
Topic 23      & \footnotesize director field, shear modulu, homeotrop, t$\lambda$, antivortex, humid, u0, hydrophil, shear band, shear strain  \\ \hline   
Topic 24      & \footnotesize signific role, key role, kibbl zurek, amino acid, play essenti, play domin, play crucial, play critical, play central, remain elus    \\ \hline   
Topic 25      & \footnotesize paramet $\eta$, ev2, rev c, rev lett, eq, right hand, right left, e e collide, e e annhil, f0   \\ \hline

\end{longtable}

\begin{longtable}[c]{p{0.2\linewidth} | p{0.8\linewidth}}
\caption{\textbf{Example of Phrase-Topic Distributions}.}
\label{sitab:topic_vectors}\\
\multicolumn{1}{p{.1\linewidth}|}{Term} & Topic-Embedding   
\endfirsthead
\hline
\
%
\vdots     &   \\      
Quantiz      &\footnotesize  [1, 0, 0, 0, 0 2259, 0, 0, 560, 0, 0, 882, 0, 0, 0, 0, 0, 0, 1, 0, 0, 0, 677, 0, 0]   \\ 
Quantum      &\footnotesize  [29, 0, 0, 21, 0, 4304, 1069, 4276, 0, 308, 0, 6008, 454, 46, 14920, 0, 0, 35931, 0, 1828, 0, 0, 1384, 7, 1]   \\  
Quark      &\footnotesize   [0, 0, 0, 0, 0, 0, 0, 0, 0, 0, 239, 0, 0, 0, 0, 0, 0, 0, 0, 0, 0, 0, 0, 0, 14542]   \\  
Quarkonia      &\footnotesize  [0, 0, 0, 0, 0, 0, 0, 0, 0, 0, 0, 0, 0, 0, 0, 0, 0, 0, 0, 0, 0, 0, 0, 0, 125]   \\  
Quarkonium      &\footnotesize  [0, 0, 0, 0, 0, 0, 0, 0, 0, 0, 0, 0, 0, 0, 0, 0, 0, 0, 0, 0, 0, 0, 0, 0, 299]   \\  
Quarter      &\footnotesize  [0, 0, 0, 0, 30, 0, 0, 0,  0, 0, 0, 321, 0, 0, 0, 0, 0, 0, 0, 0,  0, 0, 0, 0, 0]   \\  
\vdots     &   \\ 
Quantum Wire      &\footnotesize  [1, 0, 0, 0, 0, 0, 0, 0, 0, 0, 0, 342, 0, 0, 292, 0, 0, 23, 0, 0, 0, 0, 91, 0, 0]   \\
Quantum Zeno      &\footnotesize  [0, 0, 0, 0, 0, 0, 0, 0, 0, 0, 0, 0, 0, 0, 12, 0, 0, 31, 0, 0, 0, 0, 0, 0, 0]   \\  
Quark Antiquark      &\footnotesize  [0, 0, 0, 0, 0, 0, 0, 0, 0, 0, 0, 0, 0, 0, 0, 0, 0, 0, 0, 0, 0, 0, 0, 0, 433]   \\  
Quark Condens      &\footnotesize  [0, 0, 0, 0, 0, 0, 0, 0, 0, 0, 1, 0, 0, 0, 0, 0, 0, 0, 0, 0, 0, 0, 0, 0, 107]   \\  
Quark Decay      &\footnotesize  [[0, 0, 0, 0, 0, 0, 0, 0, 0, 0, 0, 0, 0, 0, 0, 0, 0, 0, 0, 0, 0, 0, 0, 0, 25]   \\  
\vdots     &   \\

\end{longtable}

\subsection{Innovation Example}\label{section:innovation}
\begin{figure}[h!]
\includegraphics[width=\linewidth]{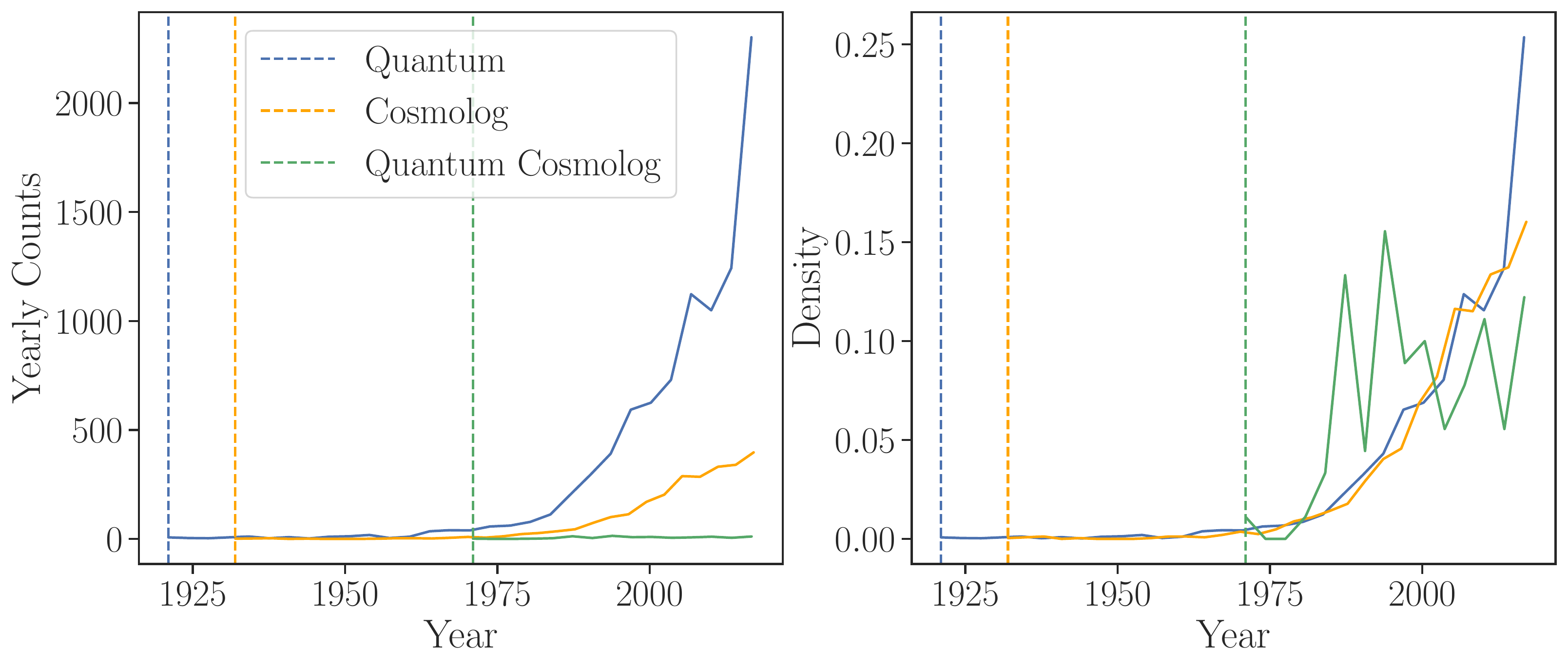}
\caption{Example of innovation measure with terms "quantum" and "cosmolog"}
\label{sifig:innovation_example}
\end{figure}

The innovation metric counts the first time a term or a new combination of terms have been seen in an article over the entire corpus. Fig. \ref{sifig:innovation_example} shows the introduction of the terms "quantum" and "cosmolog" in the corpus. Note that "cosmolog" is the root of words such as "cosmology" and "cosmological" that were lemmatized in pre-processing. We plot the frequency of the terms in time as well as vertical lines representing the first year the term has been seen. We also plot the counts of the phrase "quantum cosmolog" which is an additionally considered term in our topic model.

\clearpage
\section{Trends of Novelty and Innovation}

\begin{figure}[h!]
\includegraphics[width=\linewidth]{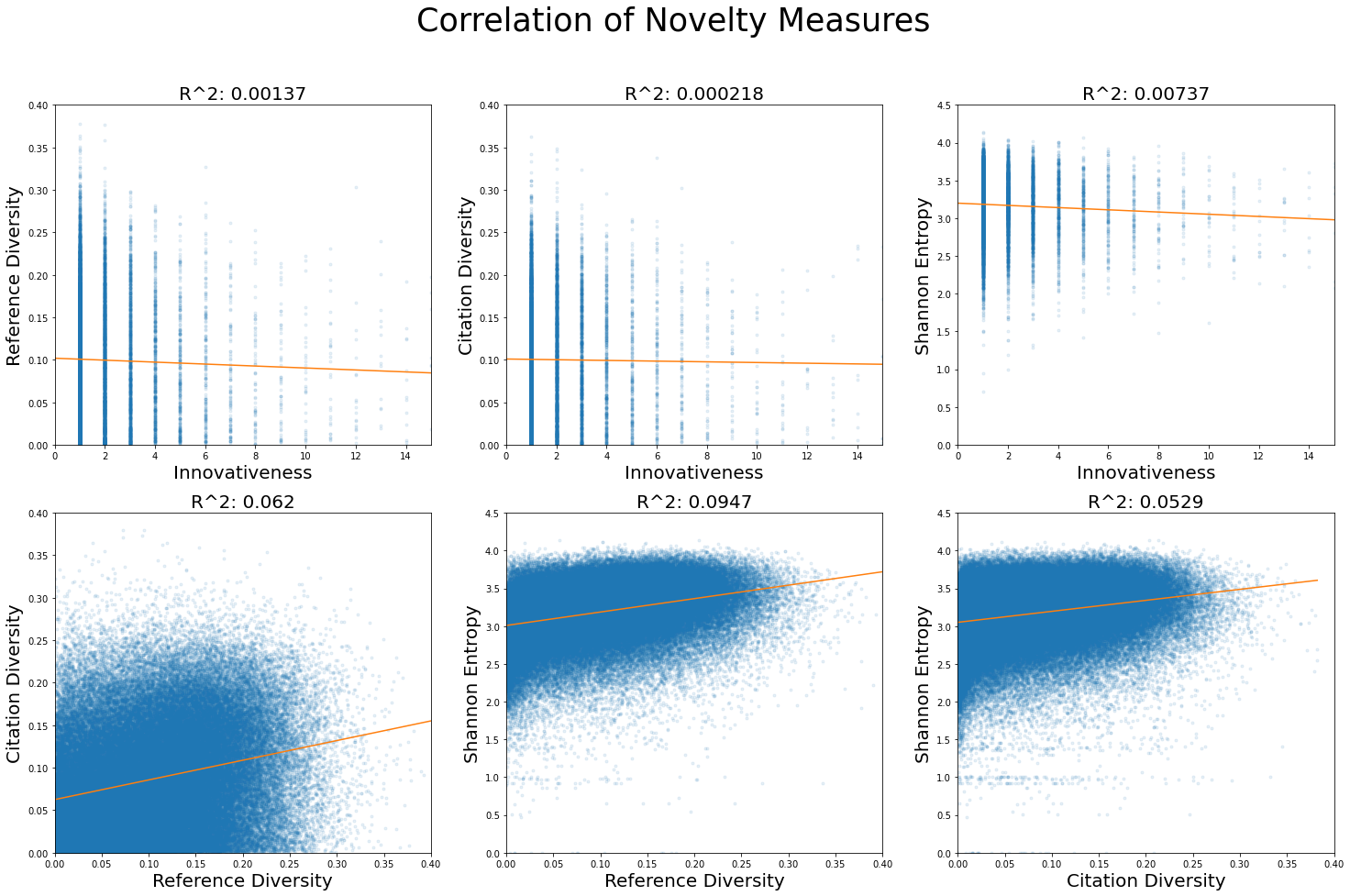}
\caption{Correlations between all novelty and innovation measures based on Pearson's $r$.}
\label{sifig:novelty_correlations}
\end{figure}

\clearpage
\section{Author and Paper-Level Novelty and Output}

\begin{figure}[h!]
\includegraphics[width=\linewidth]{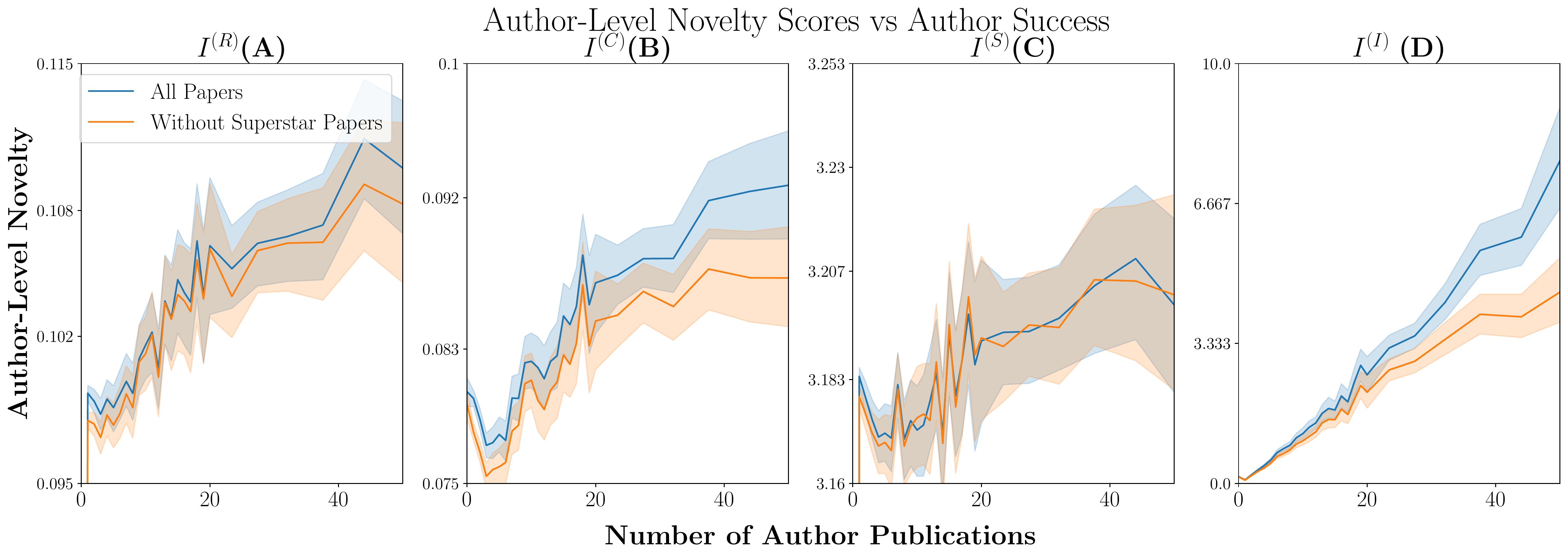}
\caption{Novelty and innovation metrics of an author's publication record as a function of their number of publications. Authors with between 1-50 publications in the corpus have been considered.}
\label{sifig:author_nov_sucess}
\end{figure}

\begin{figure}[h!]
\includegraphics[width=\linewidth]{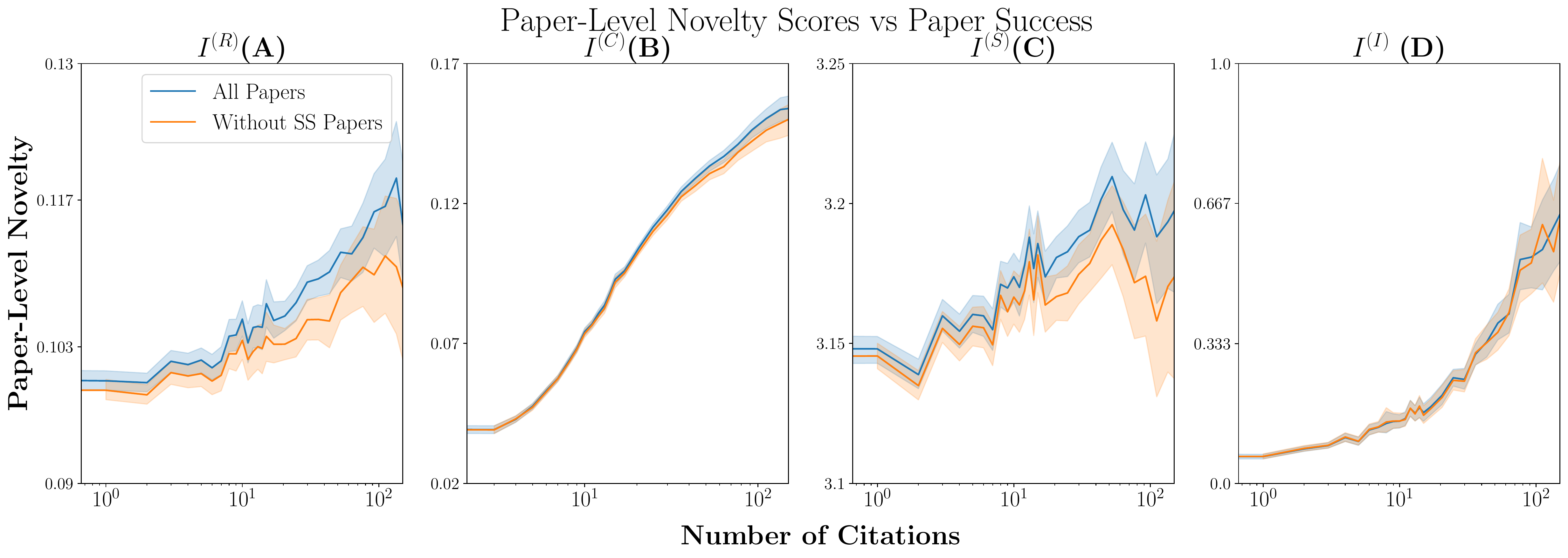}
\caption{Novelty and innovation metrics of an author's publication record as a function of the number of citations their papers garner.}
\label{sifig:paper_nov_sucess}
\end{figure}

\begin{figure}[h!]
\includegraphics[width=\linewidth]{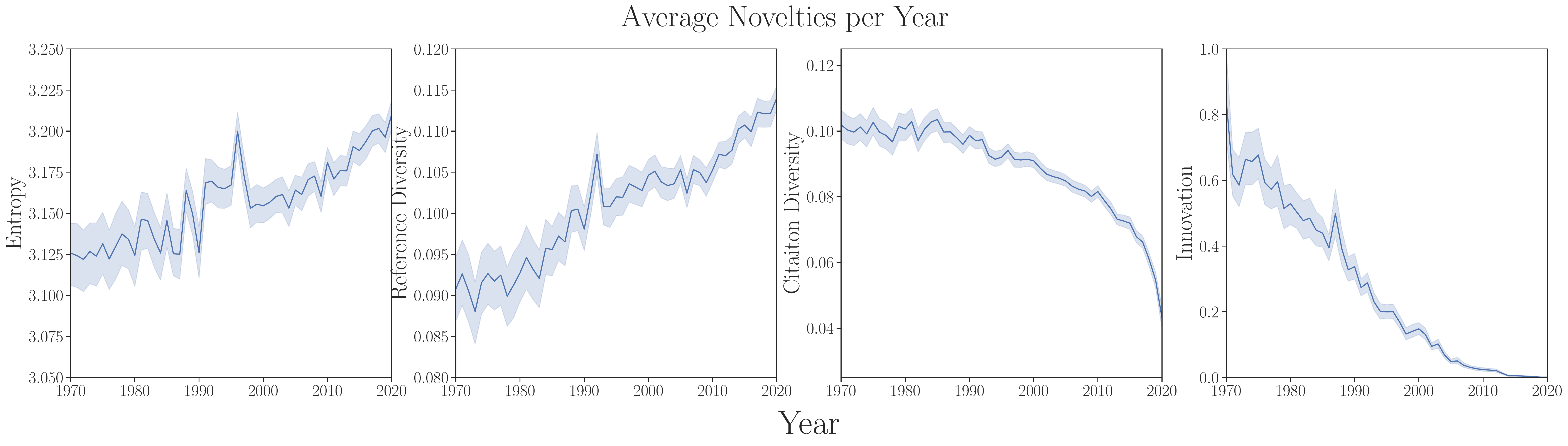}
\caption{All Novelty Measures per Year.}
\label{sifig:novelty_year}
\end{figure}

\begin{figure}[h!]
\includegraphics[width=\linewidth]{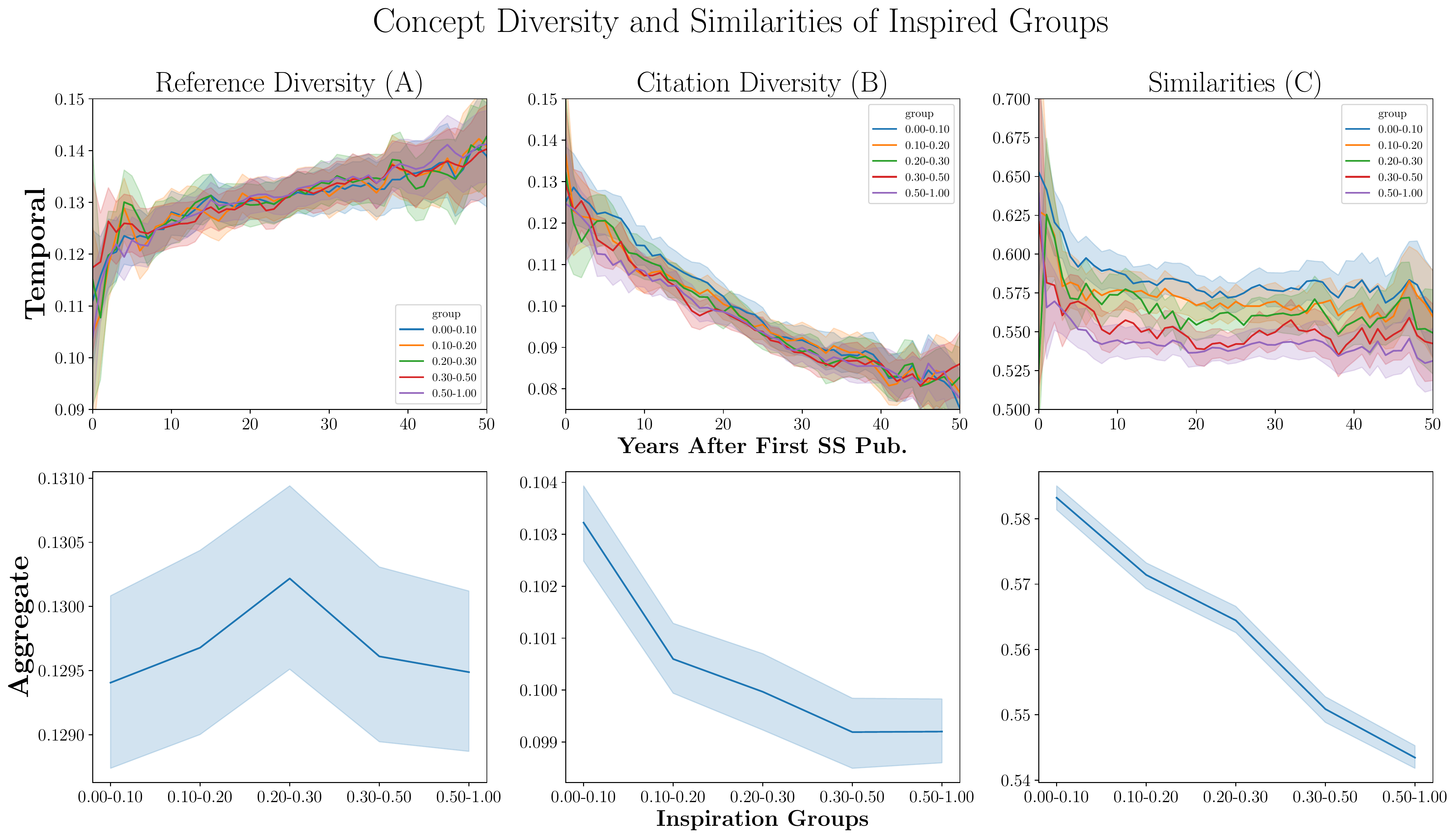}
\caption{(A) Reference Diversity, (B) Citation Diversity, (C) Within-group paper similarities for the followers of a superstar partitioned by level of inspiration. Upper panel: temporal evolution. Lower panel: averaged in time.}
\label{sifig:inspired_group_novelty_SI}
\end{figure}

\clearpage

\begin{figure}
\includegraphics[width=\linewidth]{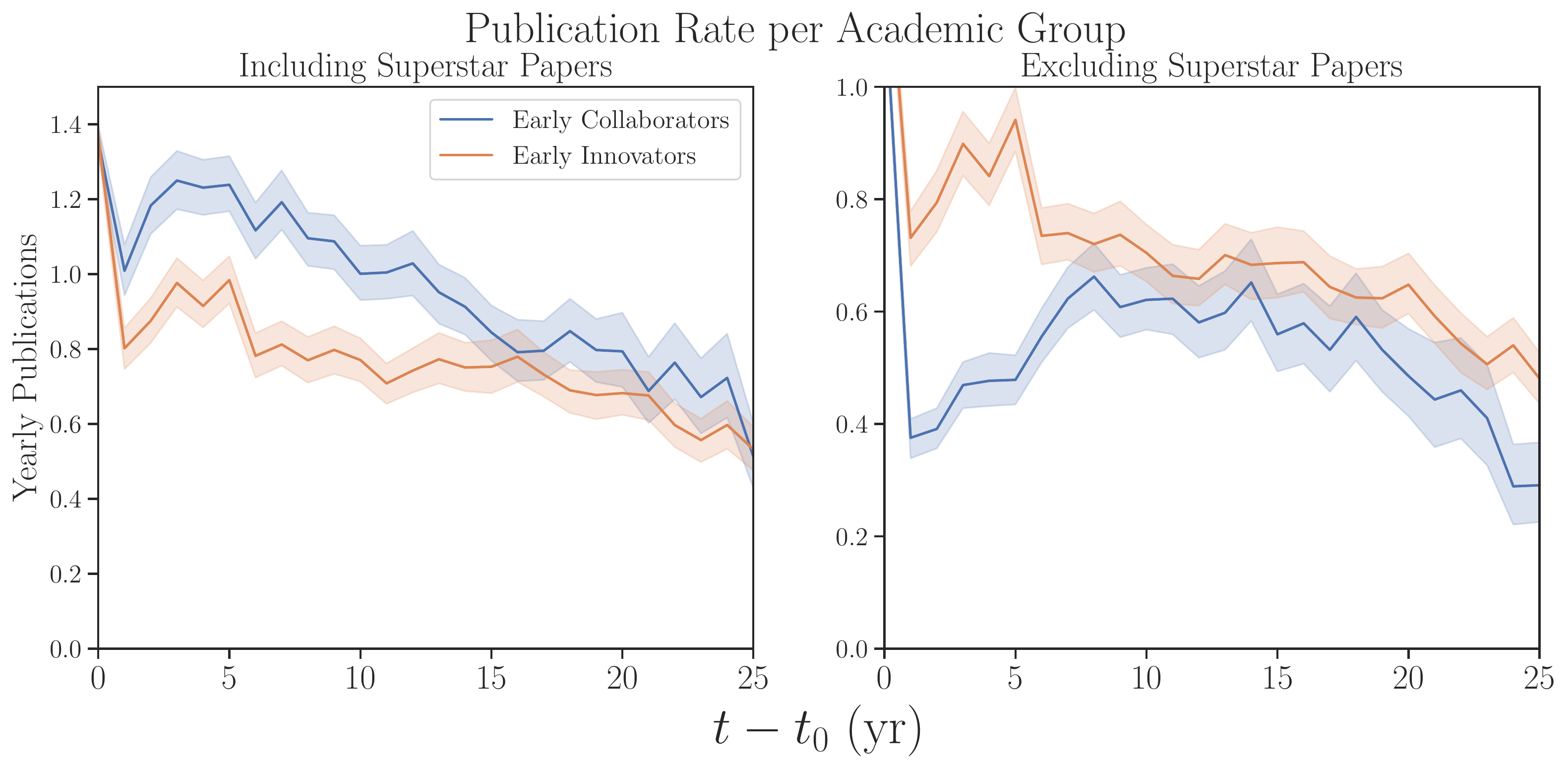}
\caption{Publication rates of academic groups, \textbf{LEFT} including superstar collabortions and \textbf{RIGHT} excluding superstar collaborations}
\label{sifig:pub_rate_academic_groups}
\end{figure}